\documentclass[10pt,twocolumn]{article}

\usepackage{isqcmc2023}
\usepackage{graphicx,url}
\usepackage{xcolor}
\usepackage{physics}
\usepackage{braket}
\usepackage{tabularray}
\usepackage{soul}
\usepackage{qcircuit}
\usepackage{cancel}
\usepackage{caption}
\usepackage{subcaption}
\usepackage[utf8]{inputenc}

\usepackage{orcidlink}
\title{Variational Quantum Harmonizer: Generating Chord Progressions and Other Sonification Methods with the VQE Algorithm}

\author{Paulo Vitor Itaboraí\inst{1}\orcidlink{0000-0002-4956-2958} \and
        Tim Schwägerl\inst{2}$^,$\inst{4} \and María Aguado Yáñez\inst{3}\orcidlink{0009-0003-7133-4681} \and Arianna Crippa\inst{2}$^,$\inst{4}\orcidlink{0000-0003-2376-5682} \and\\ Karl Jansen\inst{2}\orcidlink{0000-0002-1574-7591} \and Eduardo R. Miranda\inst{1}\orcidlink{0000-0002-8306-9585} \and Peter Thomas \inst{1}\orcidlink{0000-0001-6531-8675}  }

\address{Interdisciplinary Centre for Computer Music Research -- University of Plymouth \\
        Drake Circus - PL48AA - Plymouth - United Kingdom
         \nextinstitute
         Centre for Quantum Technologies and Applications -- Deutsches Elektronen-Synchrotron DESY\\
         Platanenallee 6 - 15738 - Zeuthen - Germany
         \nextinstitute
         Music Technology Group -- Pompeu Fabra University\\
         Roc Boronat, 138 - 08018 - Barcelona - Spain
        \nextinstitute
         Instit\"ut für Physik, Humboldt-Universit\"at zu Berlin,
         Newtonstr. 15, 12489 Berlin, Germany
         \email{Corresponding Authors: paulo.itaborai@plymouth.ac.uk, tim.schwaegerl@desy.de}
}

\begin{document}

\maketitle

\begin{abstract}

\color{black}
This work investigates a case study of using physical-based sonification of Quadratic Unconstrained Binary Optimization (QUBO) problems, optimized by the Variational Quantum Eigensolver (VQE) algorithm. The VQE approximates the solution of the problem by using an iterative loop between the quantum computer and a classical optimization routine. This work explores the intermediary statevectors found in each VQE iteration as the means of sonifying the optimization process itself. The implementation was realised in the form of a musical interface prototype named Variational Quantum Harmonizer (VQH), providing potential design strategies for musical applications, focusing on chords, chord progressions, and arpeggios. The VQH can be used both to enhance data visualization or to create artistic pieces. The methodology is also relevant in terms of how an artist would gain intuition towards achieving a desired musical sound by carefully designing QUBO cost functions. Flexible mapping strategies could supply a broad portfolio of sounds for QUBO and quantum-inspired musical compositions, as demonstrated in a case study composition, "Dependent Origination" by Peter Thomas and Paulo Itaborai.

\end{abstract}

\section{Introduction}\label{sec:intro}

There are several attempts to define what a Musical Instrument is. When considering acoustic instruments from a more physical standpoint, Fletcher and Rossing state the following.

\textit{“In most instruments, sound production depends upon the collective
behavior of several vibrators, which may be weakly or strongly coupled
together. This coupling, along with nonlinear feedback, may cause the instrument as a whole to behave as a complex vibrating system, even though the individual elements are relatively simple vibrators.” }\cite{fletcher2012physics}

The interesting point being made here is that complex behaviour could arise from simple coupled systems, which when applied to sound could provide music expressivity.

\begin{figure}[ht!]
    \centering
    \includegraphics[width=.45\textwidth]{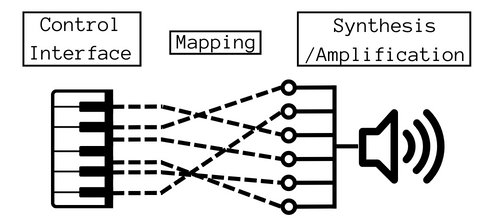}
    \caption{Constituting elements of a digital musical instrument \cite{miranda2006new}}
    \label{fig:dmi}
\end{figure}

Beyond acoustic media and mechanical vibrating systems, electronic and digital technologies have also become musical instruments. In fact, the flexibility of current music programming languages (such as SuperCollider\cite{SuperCollider}, Pure Data\cite{puckettePureDataPd}, Max/MSP\cite{Max2021}, etc.) allows the adventurous to design and build hundreds of new musical interfaces. In programming environments, performers can build, reform, and play musical interfaces in the time span of a performance \cite{roads1996computer}. Regarding this new musical interfacing ecosystem, Miranda and Wanderley \cite{miranda2006new} arrive at a concise representation model to visualise the fundamental constituting elements of a generic Digital Musical Instrument (Fig. \ref{fig:dmi}). 

According to this representation, a typical Digital Musical Instrument is consisted of three parts. First, there is a Control Interface from which the performer interacts with the instrument and expresses musical ideas. Second, there is a mapping process, in which the interface parameters are transformed and combined into control parameters. Finally, the control parameters drive a synthesis stage in which sound is generated and amplified.

It is possible to extrapolate this definition to include quantum technologies, allowing the hybridisation of quantum and classical media at the controlling and mapping layers of musical interfaces \cite{mirandaQuantumComputerMusic2022}. It is arguable that the introduction of emerging technologies in music and artistic experimentation has often led to new forms or musical expression \cite{fasciani202120}. Similarly, Quantum Computing can shed new light on quantum-inspired art and quantum aesthetics \cite{mirandaQuantumComputerMusic2022}\cite{mirandaQuantumComputingArts2022}. Also, music has historical contributions to scientific investigation \cite{maor2020music} - and it should be no different with quantum computing.

Although quantum processing units deal with quantum information and apply quantum logical operations (quantum gates), they are strongly reliant on classical computation. For example, Noisy Intermediate Scale Quantum Computers (NISQ)\cite{Preskill2018quantumcomputingin} are currently available on cloud services, and the respective instructions and algorithms are written in and controlled from classical computers. Additionally, the results from the quantum computation are also analysed and post-processed in classical computers.

In this work, we introduce the Variational Quantum Harmonizer (VQH) musical interface. It enables composition in terms of a Quadratic Unconstrained Binary Optimization (QUBO) problem and utilises the Variational Quantum Eigensolver (VQE) algorithm to solve the QUBO. The paper is structured as follows: In section \ref{(sec:vqh} we explain QUBO and how to solve it using the VQE algorithm. In this context, we introduce quantum circuits and classical optimization routines, the main constituents of the hybrid quantum-classical VQE algorithm. In section \ref{sec:sonifying}, we illustrate how to encode chords in the solution of QUBO and define a sonification of the VQE's optimization process. In section \ref{sec:qubo_ising}, we elaborate on the connection between QUBO and the mathematically equivalent Ising Hamiltonian. This enables an alternative approach to composing with VQH and the use of the Ising model to extend QUBO. In section \ref{sec:interface}, we explain the VQH's user interface and in section \ref{sec:mapping} alternative mapping strategies for sonification are presented. The possibility of using VQH in live performances is assessed in section \ref{sec:live}. We conclude and propose ways to extend VQH in section \ref{sec:conclusion}. We assume that the reader has basic knowledge of quantum computing, especially the gate-based model of computations.

\section{The Variational Quantum Harmonizer algorithm}
\label{(sec:vqh}
The \textit{Variational Quantum Harmonizer} prototype is designed as a sonification-based musical instrument. Typically, the control interface of this class of instrument contains a dataset or database obtained from a scientific experiment or a simulation. The nature of the data is often related to a poetic, aesthetic or artistic motivation of the instrument designer. This data is customarily transformed directly into a control signal, that could be used to modulate a synthesis parameter of choice or generate symbolic musical notation. 
However, it is also true that the mapping stage could use data from different origins and contain multiple layers of smaller mappings and data processing before it becomes a control signal. This usually increases the complexity and expressivity of the musical gestures.
In simple terms, the VQH control interface has two main layers. It does not contain a readily available dataset, as usual. Instead, the user \textit{designs the experiment itself}, with the aim of achieving an approximate musical result (e.g., chords, chord progressions, and other compositional gestures). The problem is then solved with a variational method using a hybrid quantum-classical algorithm, creating a respective dataset. Finally, the dataset is sonified.

Composing with the VQH works as follows: First, the user chooses the coefficients of a square matrix that defines a QUBO problem (Sec. \ref{sec:qubo}. QUBO is a common formulation of a variety of optimization problems from Logistics to Machine Learning. Then, the VQH sonifies the process of solving the QUBO. One could, for example, construct a QUBO such that its solution leads to a simple chord. To solve the QUBO on a gate-based quantum computer using the Variational Quantum Eigensolver (VQE), it is transformed to a mathematical equivalent Ising Hamiltonian. Details on VQE and on the transformation from QUBO to an Ising Hamiltonian are found in \ref{sec:vqe}. Then, VQE optimizes the parameters of a quantum circuit to approximate the ground state of the Hamiltonian and therewith the QUBO problem using an iterative loop between a quantum computer and a classical optimization routine. The results obtained during the quantum algorithmic process are analyzed, post-processed and finally used as sonification data to drive various synthesis parameters, as specified by the artist.

\subsection{Quadratic Unconstrained Binary Optimization}
\label{sec:qubo}

The Quadratic Unconstrained Binary Optimization problem can be defined in terms of a cost function (eq. \ref{eq:qubo}) that contains linear and quadratic terms. Individual agents ($n_i \in \{0,1\}$) have both an individual contribution to the system (linear coefficients $a_i$), as well as interfere with other agents (quadratic coefficients $b_{ij}$). Solving this problem means finding (or approximating) a configuration that \textit{ minimizes} the cost function.

\begin{equation}
    Q(n) = \sum_i^N a_i n_i + \sum_{i,j}b_{i,j}n_in_j; \quad n \in \{0,1\}^N
    \label{eq:qubo}
\end{equation}

Currently, quantum computing approaches are being investigated for solving QUBO, for example, in Quantum Machine Learning, using the VQE algorithm \cite{date2021qubo}. 

\subsection{The Variational Quantum Eigensolver}
\label{sec:vqe}

VQE is a direct application of the variational method used in quantum mechanics. It was proposed in 2014 by Peruzzo \cite{Peruzzo2014} to approximate the ground state energy of molecules. The expectation value of a Hamiltonian $H$ for a state $| \psi \rangle$ may be expressed as a weighted sum of its eigenvalues $\lambda_i$, where $| \psi_i \rangle$ are the corresponding eigenvectors:
\begin{align}
    \langle H \rangle_{\psi} \equiv \langle \psi | H | \psi \rangle &=  \sum_{i = 1}^{N} \lambda_i | \langle \psi_i | \psi \rangle |^2
\end{align}
Since $| \langle \psi_i | \psi\rangle |^2 \ge 0 $, the expectation value will always be at least the minimum eigenvalue:
\begin{align}
    \lambda_{\text{min}} \le \langle H \rangle_{\psi}
\end{align}
By applying a parameterized circuit $U(\vec{\vartheta})$ to some arbitrary starting state $|\psi\rangle$, VQE provides an estimate $\lambda_{\vec{\vartheta}}$ bounding $\lambda_{\text{min}}$:
\begin{align}
    \lambda_{\text{min}} \le \lambda_{\vec{\vartheta}} \equiv \langle \psi(\vec{\vartheta)} |H|\psi(\vec{\vartheta}) \rangle \\
    \text{with} \quad U(\vec{\vartheta})|\psi\rangle \equiv |\psi(\vec{\vartheta})\rangle \nonumber
\end{align}
The estimate is then iteratively optimized by a classical optimizer updating the parameters $\vec{\vartheta}$. In practice the expectation value $\lambda_{\vec{\vartheta}}$ is approximated as the sample mean of repeatedly preparing $|\psi(\vec{\vartheta})\rangle$ and measuring $H$ in this state $K$ times:
\begin{align}
    \lambda_{\vec{\vartheta}} \approx \frac{1}{K} \sum_{k=1}^K \lambda_{\vec{\vartheta},k}
\end{align}

When using VQE to solve QUBO one transforms the function $Q(T)$ to a computational equivalent Ising Hamiltonian by mapping the binary variables $n_i$ to Pauli $Z_i$ operators: 
\begin{align}
H(Z)=\sum_i^N a_i^{VQE} Z_i + \sum_i^N\sum_{j<i}^N b_{ij}^{VQE}Z_i Z_j
\label{eq:qubotoqubit}
\end{align}

The difference between binary variables and the eigenvalues of the Pauli $Z_i$ operators can be accounted for by transforming the coefficients:
\begin{align}
    n_i \Rightarrow Z_i = 1-2n_i \\
    b_{ij} \Rightarrow b_{ij}^{VQE} = b_{ij} \\ 
    a_i \Rightarrow a_i^{VQE} = -2a_i - \sum_{j\ne i}^{N}b_{ij}
    \label{eq:qubotoising_transform}
\end{align}

\subsection{Parameterized Quantum Circuits}
\label{sec:circuits}
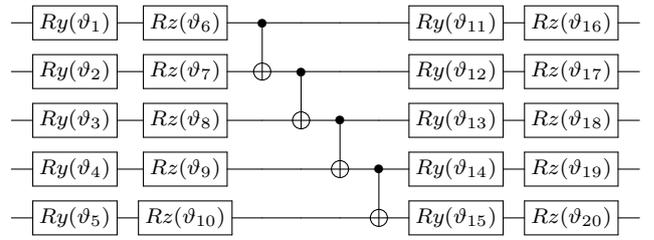
\begin{figure}
    \centering
    \footnotesize
\Qcircuit @C=1em @R=.7em {
& \gate{Ry(\vartheta_1)} & \gate{Rz(\vartheta_{6})} & \ctrl{1} & \qw & \qw & \qw & \gate{Ry(\vartheta_{11})} & \gate{Rz(\vartheta_{16})} & \qw \\
& \gate{Ry(\vartheta_2)} & \gate{Rz(\vartheta_{7})} & \targ & \ctrl{1} & \qw & \qw & \gate{Ry(\vartheta_{12})} & \gate{Rz(\vartheta_{17})} & \qw \\
& \gate{Ry(\vartheta_3)} & \gate{Rz(\vartheta_{8})} & \qw & \targ & \ctrl{1} & \qw & \gate{Ry(\vartheta_{13})} & \gate{Rz(\vartheta_{18})} & \qw \\
& \gate{Ry(\vartheta_4)} & \gate{Rz(\vartheta_{9})} & \qw & \qw & \targ & \ctrl{1} & \gate{Ry(\vartheta_{14})} & \gate{Rz(\vartheta_{19})} & \qw \\
& \gate{Ry(\vartheta_5)} & \gate{Rz(\vartheta_{10})} & \qw & \qw & \qw & \targ & \gate{Ry(\vartheta_{15})} & \gate{Rz(\vartheta_{20})} & \qw \\
}
    \caption{An EfficientSU2 circuit for 5 qubits. The initial $R_y$ rotations are referred to as the 0-thy layer. The 1st layer here indroduces linear entanglement using $CNOT$ gates and additional rotations.}
    \label{fig:circuit}
\end{figure}

As explained in the previous section, a main constituent of the VQE algorithm is a parameterized circuit $U(\vec{\vartheta})$ that prepares a trial state $|\psi(\vec{\vartheta})\rangle$. There are two paradigms to design the circuit, problem specific circuits and hardware efficient circuits. While the former encodes knowledge of the problem into the circuit, the latter uses alternating layers of rotation gates and entangling gates. This structure results in shallow circuits that can be executed on NISQ computers. One prominent example is the \texttt{EfficientSU2} circuit, displayed in figure \ref{fig:circuit} for 5 qubits. The initial layer of $R_y$ rotations is sufficient to express the solution of QUBO, because it is a computational basis states. However, introducing additional layers of entangling $CNOT$ gates and $Ry$ gates enables the circuit to reach states that lie in larger Hilbert spaces. The user can specify the number of layers and the entanglement structure to produce different sounds. The VQH implementation in this work has used \texttt{EfficientSU2} circuits with two layers (the initial layer and one entanglement layer).

\subsection{Classical Optimizers}
\label{sec:optimizers}
Another key ingredient of the VQH algorithm is the choice of the classical optimization routine. From a conceptual point of view, one differs gradient-based and gradient free optimizers. For our study, however, the most audible effect is due to how the optimizer updates the parameters. For instance, SPSA \cite{spsa1992} perturbates all parameters at once, whereas optimizers such as NFT \cite{nakanishi2020sequential} change only one parameter at a time. VQH supports all optimizers in Qiskit \cite{Qiskit}. For explanations on how these optimizers work, we refer to Qiskit and the original implementation of many optimizers in Scipy \cite{2020SciPy-NMeth}. 

\section{Sonifying the VQE algorithm: Chords and Chord progressions}
\label{sec:sonifying}

In a previous work conducted by Clemente, Crippa, Jansen and Tüysüz \cite{clemente2022new}, the implemented mapping strategy is to transform quantities computed with a quantum variational approach into audible frequencies. 
In this work, they considered the Ising model and studied the Hamiltonian ground state with VQE and excited states with the Variational Quantum Deflation (VQD)\cite{PhysRevA.99.062304,Higgott2019variationalquantum} algorithm.
The first approach is to apply the VQD algorithm, compute the energy eigenvalues and convert them to audible frequencies. Using the coupling $h_x$ as a ‘\textit{time variable}' (see eq. \ref{eq:isingtransverse} at the conclusion), they followed the behaviour of the corresponding frequencies and play
them through an output device. 
They also proposed to measure the observables during the optimization itself and thus play sounds from variational algorithms.
In the last part, they followed again the approach of assigning the role of ‘time variable’ to the external magnetic field. Frequencies can then be computed from the energy eigenvalues and intensities from the magnetization measured in the corresponding eigenstates.

In contrast, this study adopts a different perspective on the mapping strategy, focusing the attention on the variational method itself, as well as emphasising the quantum states associated with the computed energies. Specifically, our approach incorporates the statevector sonification of each iteration of the optimization process, as detailed in this section. As a result, this method may enable an unique auditory insight of the intricate interplay between different classical optimizers and the way they navigate and reach a QUBO solution when assisted by a quantum algorithm.

For instructive purposes, the mapping strategy will be explained in this section using a simple example.

\subsection{Qubit words as chords: the QUBO Harp}
\label{sec:qubit_chords}

Imagine a special \textit{ harp} with 12 "binary" strings. Each string is tuned for a specific frequency, e.g. 1 octave of the chromatic scale. A convention was established where $0$ represents a resting string - or \textit{silence} - while $1$ reciprocally indicates a \textit{sounding note}. As a result, a 12-digit binary word can be used to represent notes, intervals, or more generally, \textit{Chords} (For simplicity, all cases will be referred to as "chords" for the remaining of this text). For instance, a C Major chord could be represented as shown below.

\begin{figure}[ht]
    \centering
    \begin{tabular}{cccccccccccc}

        $1$\, &  $0$\, & $0$\, & $0$\, & $1$\, & $0$\, & $0$\, & $1$\, & $0$\, & $0$\, & $0$\, & $0$\,
    \\

    \end{tabular}
    
    \begin{tabular}{cccccccccccc}
    \hline
     \scriptsize{$C$}\!  & \!\scriptsize{$C\#$}\! & \!\scriptsize{$D$}\! & \!\scriptsize{$D\#$}\!\! & \!\scriptsize{$E$} & \scriptsize{$F$}\! & 
     \!\scriptsize{$F\#$}\! & \!\scriptsize{$G$}\! & \!\scriptsize{$G\#$}\! & \!\scriptsize{$A$}\! & \!\scriptsize{$A\#$}\! & \!\scriptsize{$B$}
     \\\hline
    \end{tabular}
    \caption{A C Major Chord represented in binary}
    \label{fig:binaryharp}
\end{figure}

However, the strings of this figurative harp can also be \textit{damped} and/or \textit{coupled} together, leading to some kind of constructive or destructive interference that can be described as coefficients of eq. \ref{eq:qubo}. A cost function is designed, assuming that a sounding CMajor chord is a configuration that minimizes the QUBO function $Q(n)$.

This QUBO then is mapped to an equivalent Hamiltonian as in eq. \ref{eq:qubotoqubit}. In other words, we assign $\ket0$ to silence and $\ket1$ to a sounding note, and have $\ket{Cmaj}$ be the system's \textit{ground state}. 

\subsubsection{Using the VQE to minimize (and sonify) the QUBO Problem}
\label{sec:dataset}
After designing the operator coefficients, the VQE algorithm can be used to navigate the state space and approximate the solution using a hardware-efficient SU(2) ansatz (Fig. \ref{fig:circuit}). Evidently, in this case the expected result for the minimum energy eigenstate would be $\ket{Cmaj}$, \textit{as designed}. This apparently did not provide any new information apart from a floating point number representing a step-by-step estimation of the expectation value for the ground state energy $\bra{Cmaj}H\ket{Cmaj}$. However, the intention of this process is to investigate \textit{how} the result was approached.

Consequently, in our methodology, at each optimization step of the VQE, the current statevector is being sampled, stored, and used as input of the next iteration. In other words, our approach is to keep track of how the quantum state is being transformed as the algorithm looks for an optimal solution. 

\textit{The sampled state vectors of each VQE iteration (in addition to the corresponding partial estimation results of the expectation value) represents the data used for sonification.}

\subsubsection{The Statevector sonification}
The main idea of this approach is to focus on the sonification of the sampled statevectors, conveying a sounding chord where each qubit in the $\ket1$ state represents a playing note. However, since qubits may be in superposition states, note that a Cmajor-sounding chord can be achieved with many different quantum states. For example, $\ket{Cmaj_1}$, $\ket{Cmaj_2}$ and $\ket{Cmaj_3}$ below will all \textit{sound} as a C major to our hypothetic harp.

\begin{multline}
  \;\;\; \ket{Cmaj_1} = \ket{100010010000} \\\\
  \ket{Cmaj_2} = \frac{1}{\sqrt2}\left[\ket{100010000000} + \ket{000010010000}\right] \\\\
  \ket{Cmaj_3} = \sqrt{\frac{3}{4}}\ket{100000000000} + \sqrt{\frac{3}{16}}\ket{000010000000} \\+ \sqrt{\frac{1}{16}}\ket{100000010000}
\end{multline}

As a result, all possible states in which a note is playing need could be somehow grouped together.
To achieve this, the Marginal Distribution of each sampled statevector was obtained. The Marginal Distribution refers to the probability of each qubit being in the state $\ket{1}$ independently of the other qubits. By collecting the coefficients of each state contributing to a specific note to be playing, we introduce a notion of a note's relative \textit{loudness} within that chord. As a result, an \textit{additive synthesis} perspective is achieved and taken as the initial strategy to sonify the 12 coefficients of the marginal distribution.

\subsection{Example 1: Linear Cmajor Chord}

As an initial example, the CMajor cost function below is designed to use only linear coefficients \ref{eq:qubocmaj}. The desired notes (C, E, G) contribute to lower the cost, whereas the remaining notes give energy penalties when they are played. In our harp analogy, this would mean to \textit{damp} unwanted notes while favouring the resonance of preferred notes. 

\begin{equation}
    Q(n_{Cmaj}) = - n_C - n_E-n_G+\sum_{k\notin Cmaj} n_k
    \label{eq:qubocmaj}
\end{equation}

\begin{figure}[ht!]
    \centering
    \includegraphics[width=.45\textwidth]{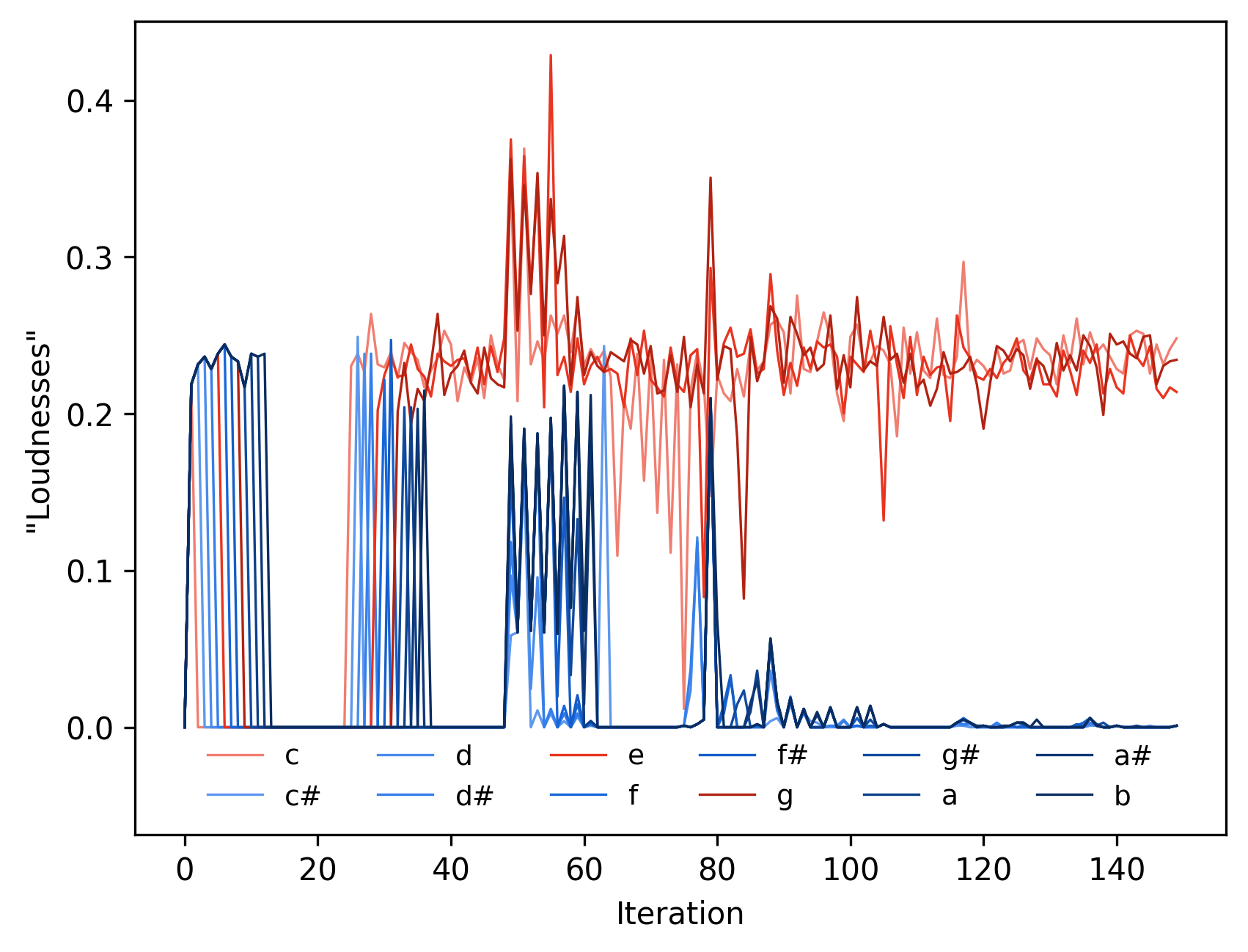}
    \caption{Marginal Distribution Evolution for Example 1}
    \label{fig:cmajorlinear}
\end{figure}

By running the VQE - using the COBYLA optimizer - to solve the problem above, the marginal distribution evolution obtained over 150 iterations is depicted in fig. \ref{fig:cmajorlinear}.

\begin{figure}[ht!]
    \centering
    \includegraphics[width=.45\textwidth]{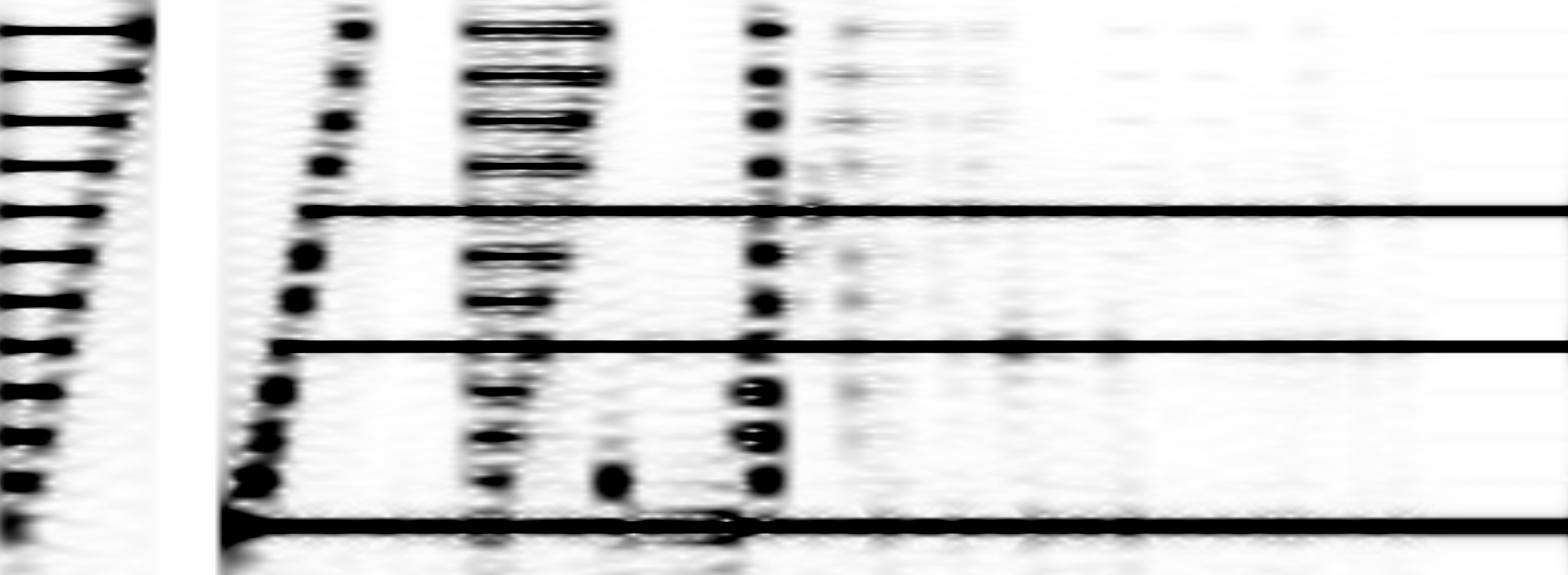}
    \caption{Visualization of Example 1 sonified with additive synthesis.}
    \label{fig:cmajorlinearspec}
\end{figure}

It is possible to observe how the COBYLA optimizer handled the minimization process, and how it guided the configuration space calculated by the quantum circuit. Most notably, the algorithm "perturbs" each note individually. Using the harp analogy again, it looks like the algorithm is "sweeping" through the strings (or muffling them). This can be heard objectively when an additive synthesis approach is applied to sonify fig. \ref{fig:cmajorlinear}. A spectrogram visualization of the sonified result is shown in fig \ref{fig:cmajorlinearspec}.

\subsection{Example 2: I-IV-V-I Linear Chord Progression}

The VQH Implementation also enables the concatenation of QUBOs. By inputting a list of QUBO matrices and a number of iterations, the VQH will proceed as follows: First, the initial QUBO is solved, as usual. Then, the resulting statevector is used as the initial point for the next QUBO optimization, and so on. In other words, the structure of the system \textit{changes over time}, affecting the optimal solution. The intended result is to create progressions of distinct chords, or create a more (musically) dynamic output.

\begin{figure}[ht!]
    \centering
    \includegraphics[width=.45\textwidth]{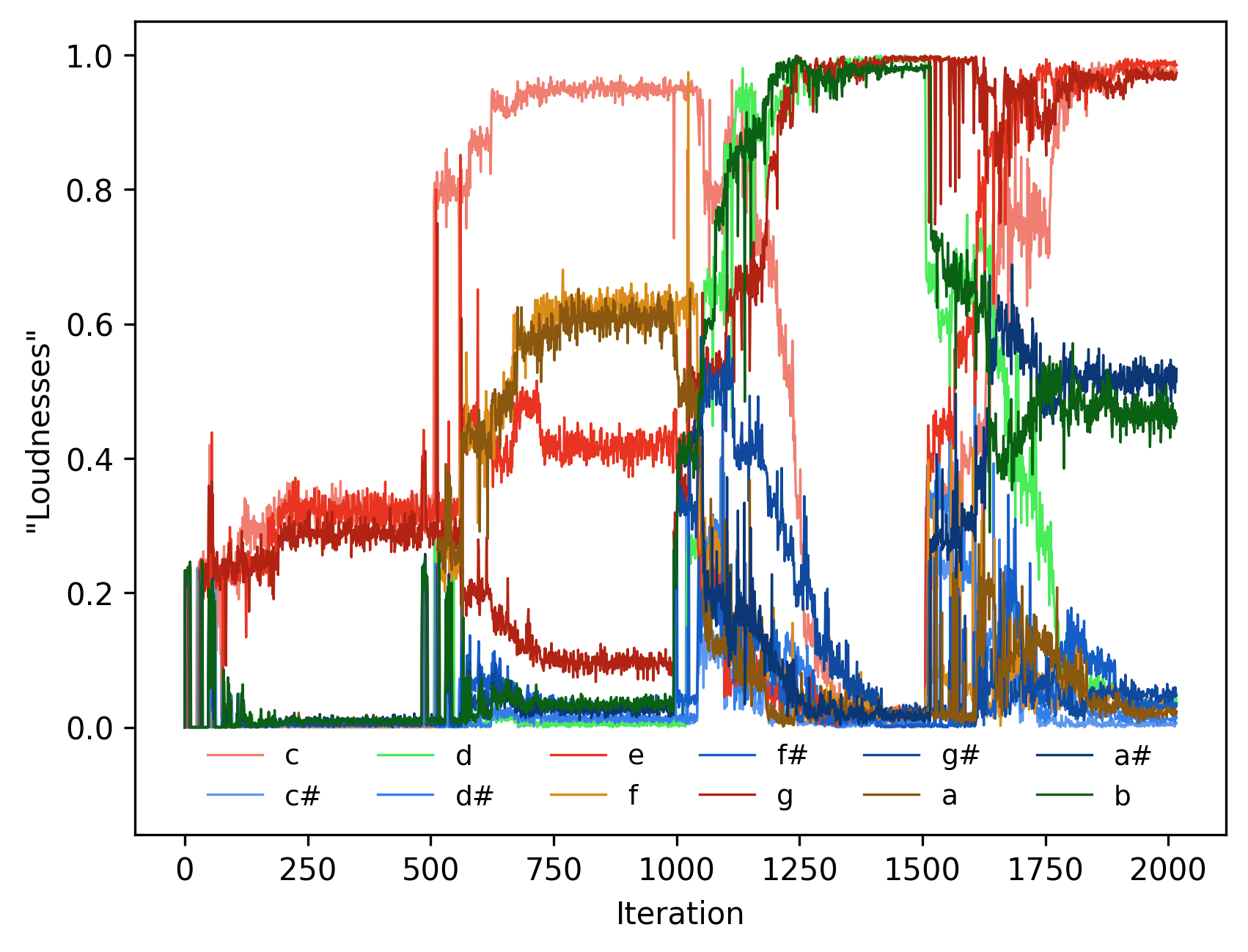}
    \caption{Marginal Distribution Evolution for Example 2}
    \label{fig:iivvilinear}
\end{figure}

\begin{figure}[ht!]
    \centering
    \includegraphics[width=.45\textwidth]{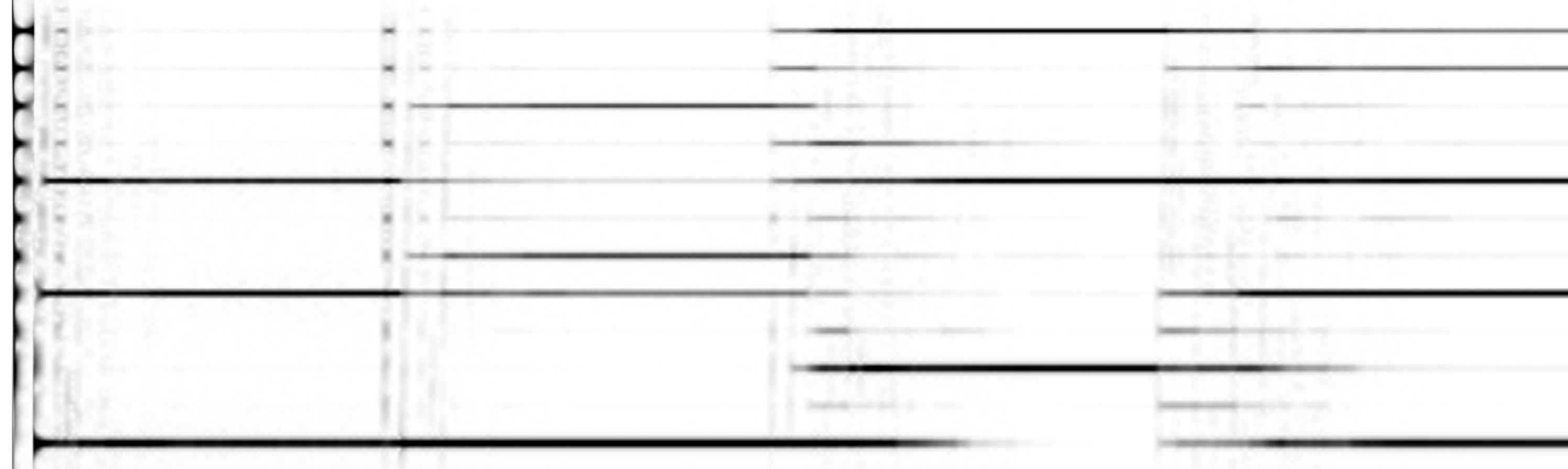}
    \caption{Visualization of Example 1 sonified with additive synthesis.}
    \label{fig:iivvilinearspec}
\end{figure}

For instance, expanding from the initial example, consider a I-IV-V-I progression. It can be defined as a set of four QUBOS (eqs \ref{eq:q1}-\ref{eq:q4}). Using the same VQE configurations as in Example 1 - using 150 iterations for each chord - leads to results as depicted in Figs. \ref{fig:iivvilinear} and \ref{fig:iivvilinearspec}.

\begin{align}
    Q_1(n^I) = - n_C - n_E-n_G+\sum_{k\notin Cmaj} n_k \label{eq:q1}\\
    Q_2(n^{IV}) = - n_F - n_A-n_C+\sum_{k\notin Fmaj} n_k\\
    Q_3(n^V) = - n_G - n_B-n_D+\sum_{k\notin Gmaj} n_k\\
    Q_4(n^I) = Q_1(n^I) \label{eq:q4}
\end{align}

Note in fig. \ref{fig:iivvilinear} how some notes persist through the progression (e.g., the E sounding in the second chord, or the B sounding in the last one), as well as completely new "\textit{intrusive}" notes appear (e.g, the A\# in the last chord). More importantly, compare the first and the last chords. Even though their QUBOS are identical, the starting point is different, leading to a new path on the configuration space, and encountering potential barren plateaus.
% \textcolor{red}{Would this  justifies the use of a quantum circuit for obtaining the statevectors (???)}  

\section{QUBOs and Ising Hamiltonians}
\label{sec:qubo_ising}
Equation \ref{eq:qubotoqubit} explicits a close structural proximity between the QUBO function and the Ising Model. It is possible to take advantage of this proximity to start framing the QUBO matrix  both as an optimization problem or as a physical system that can exhibit quantum phenomena, bringing quantum-inspired ideas to the VQH context. The \textit{harp} analogy could be reframed as a \textit{spin lattice}.

\subsection{The Ising Harp}
\begin{figure}[ht]
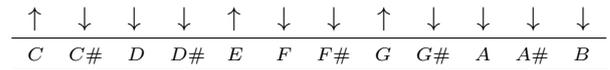

    \centering
    \begin{tabular}{cccccccccccc}

        $\uparrow$\, &  $\downarrow$\, & $\downarrow$\, & $\downarrow$\, & $\uparrow$\, & $\downarrow$\, & $\downarrow$\, & $\uparrow$\, & $\downarrow$\, & $\downarrow$\, & $\downarrow$\, & $\downarrow$
    \\

    \end{tabular}
    
    \begin{tabular}{cccccccccccc}
    \hline
     \scriptsize{$C$}\!  & \!\scriptsize{$C\#$}\! & \!\scriptsize{$D$}\! & \!\scriptsize{$D\#$}\!\! & \!\scriptsize{$E$} & \scriptsize{$F$}\! & 
     \!\scriptsize{$F\#$}\! & \!\scriptsize{$G$}\! & \!\scriptsize{$G\#$}\! & \!\scriptsize{$A$}\! & \!\scriptsize{$A\#$}\! & \!\scriptsize{$B$}
     \\\hline
    \end{tabular}
    \caption{Spin configuration of a C Major Chord}
    \label{fig:isingharp}
\end{figure}

Consider a 1D Ising problem, containing 12 spins. Then, a unique musical note is assigned to each spin in the lattice (such as the chromatic scale). By convention, a \textit{ spin down} $\downarrow$ state represents silence, and \textit{spin up} $\uparrow$ reciprocally represents a \textit{sounding note} (Fig. \ref{fig:isingharp}).

The initial strategy (Examples 1 and 2) was to achieve this desired Hamiltonian focusing on the linear part of the model, as in the QUBO function with only linear coefficients. The energy required for a desired note to be playing is favoured in relation to a specific alignment with an external longitudinal magnetic field in the Z direction, while the energies of the remaining spins are penalized, forcing them to align in the opposite direction (eq. \ref{eq:cmajor_ising}).

\begin{equation}
    H(\sigma) = h_z \Bigl(\sum_{k\in Cmaj}\sigma_k^Z - \sum_{k\notin Cmaj}\sigma_k^Z \Bigr)
    \label{eq:cmajor_ising}
\end{equation}

\subsection{Example 3: Coupled-Spin Cmajor Chord}
A second strategy would consider only the coupling between neighbouring spins, to benefit either a ferromagnetic-like ($\uparrow\uparrow$, $\downarrow\downarrow$) or anti-ferromagnetic-like ($\uparrow\downarrow$, $\downarrow\uparrow$) alignment. A QUBO function is sketched using only the quadratic terms (eq. \ref{eq:quadratic_ising_cmaj1}). Figure \ref{fig:quadraticisingcmajor} illustrates how the coefficients were designed for this example, taking into account their positivity.
\begin{figure}[ht]
    \centering
    \begin{tblr}{c|[dashed]c|[dashed]c|[dashed]c|[dashed]c|[dashed]c|[dashed]c|[dashed]c|[dashed]c|[dashed]c|[dashed]c|[dashed]c}

        $\uparrow$\, &  $\downarrow$\, & $\downarrow$\, & $\downarrow$\, & $\uparrow$\, & $\downarrow$\, & $\downarrow$\, & $\uparrow$\, & $\downarrow$\, & $\downarrow$\, & $\downarrow$\, & $\downarrow$
    \\

    \end{tblr}

    \begin{tabular}{ccccccccccc}
    % \hline
     \tiny{$>\!\!0$}  & \!\tiny{$<\!\!0$} & \!\tiny{$<\!\!0$} & \!\tiny{$>\!\!0$} & \!\tiny{$>\!\!0$} & \!\tiny{$<\!\!0$} & 
     \!\tiny{$>\!\!0$} & \!\tiny{$>\!\!0$} & \!\tiny{$<\!\!0$} & \!\tiny{$<\!\!0$} & \!\tiny{$<\!\!0$}
     \\
    \end{tabular}
    \caption{Spin configuration of a C Major Chord}
    \label{fig:quadraticisingcmajor}
\end{figure}
\begin{equation}
    Q(n^{Cmaj_{Quad}}) = \sum_{k,l}b_{kl}n_{k}n_l
    \label{eq:quadratic_ising_cmaj1}
\end{equation}
\begin{equation}
    \begin{aligned}
        b_{kl} = 
\begin{cases}
    >0, &\begin{aligned}
        & k \in Chord;\, l \notin Chord \\ & k \notin Chord;\, l \in Chord
    \end{aligned} \\
    <0, & \text{otherwise}
\end{cases}
    \end{aligned}
    \label{eq:quadratic_ising_cmaj2}
\end{equation}

\begin{figure}[ht!]
    \centering
    
    \includegraphics[width=.45\textwidth]{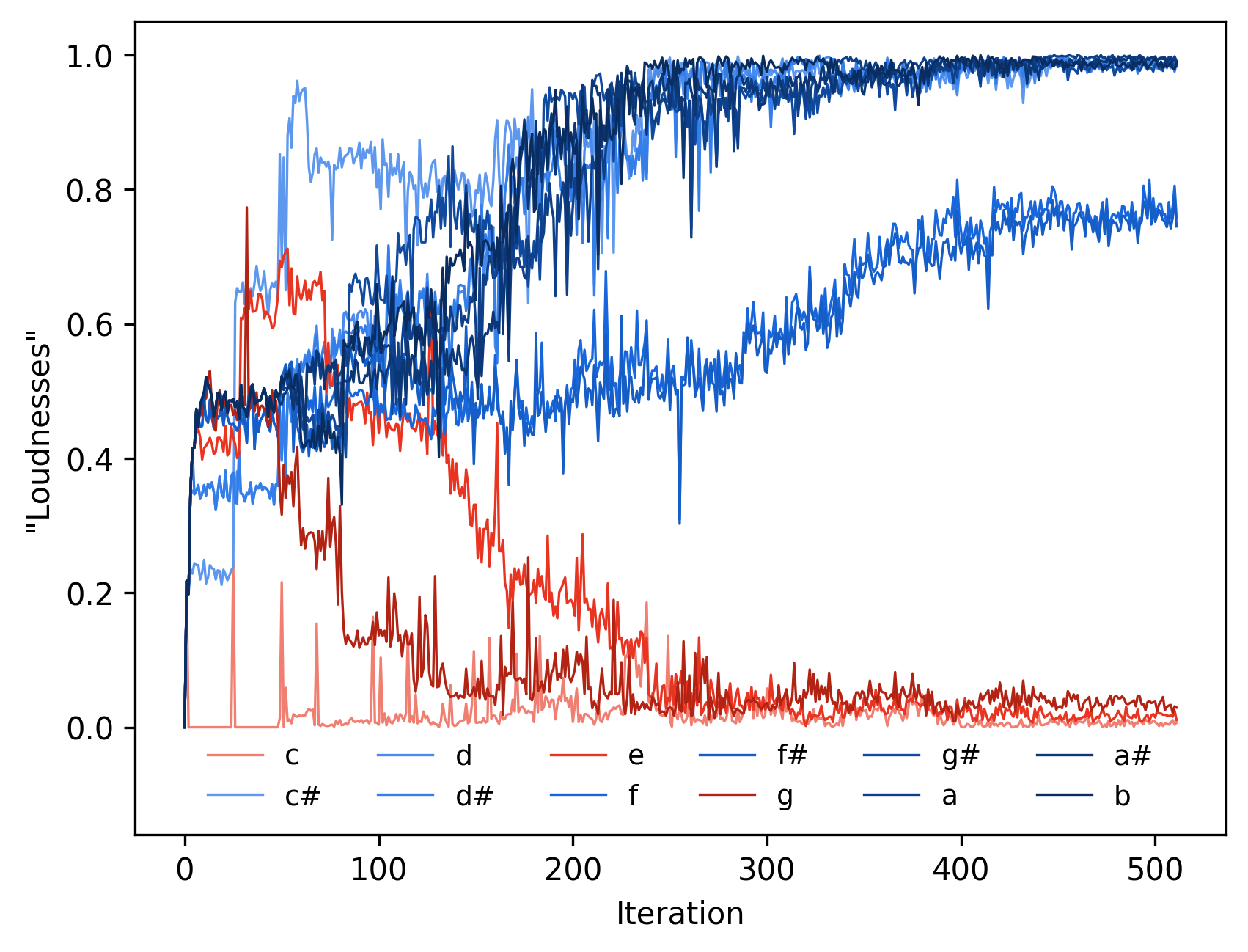}
    \caption{VQH result for Example 3}
    \label{fig:anticmaj}
\end{figure}

Note the optimization of eq. \ref{eq:quadratic_ising_cmaj1}  leads to an (apparently) unexpected solution  (Fig. \ref{fig:anticmaj}).  Moreover, the result seems to have arrived at the \textit{opposite} of the intentioned $C_{maj}$  chord, having all other notes but C, E, and G playing. 
The designed system  eq. \ref{eq:quadratic_ising_cmaj1}  focussed on spin couplings, but did not account for differences in global orientation. As a result, the state $\ket{antiC_{maj}}\ket{\downarrow\uparrow\uparrow\uparrow\downarrow\uparrow\uparrow\downarrow\uparrow\uparrow\uparrow\uparrow}$ became a solution for the problem.

In fact, it is possible to verify (using eqs. \ref{eq:qubotoqubit}-\ref{eq:qubotoising_transform}) that the originally intended chord is Not a solution to the problem, since the QUBO quadratic coefficients also impact the Ising's linear coefficients (eq. \ref{eq:qubotoising_transform}):

\begin{equation} \label{eq:expected_anticmaj}
\begin{split}
H = - \sum_i \big(\cancelto{0}{2a_i}+\sum_{i\neq j}b_{ij}\big)Z_i + \sum_{i\neq j}b_{ij}Z_iZj\\
\bra{\downarrow\uparrow\uparrow\uparrow\downarrow\uparrow\uparrow\downarrow\uparrow\uparrow\uparrow\uparrow}H\ket{\downarrow\uparrow\uparrow\uparrow\downarrow\uparrow\uparrow\downarrow\uparrow\uparrow\uparrow\uparrow} = \\ 
 = (-12) - 12 =& -24
\end{split}
\end{equation}

\begin{equation}
\begin{split}
\bra{\uparrow\downarrow\downarrow\downarrow\uparrow\downarrow\downarrow\uparrow\downarrow\downarrow\downarrow\downarrow}H\ket{\uparrow\downarrow\downarrow\downarrow\uparrow\downarrow\downarrow\uparrow\downarrow\downarrow\downarrow\downarrow} \\= (+12) -12 &= 0
\label{eq:expected_cmajor}
\end{split}
\end{equation}

\subsection{Example 4: Chords in Superposition}
In the attempt of correcting the last example, the QUBO linear coefficients ($a_i$) will be re-introduced, with the intention of closing the energy gap between $\ket{\uparrow\downarrow\downarrow\downarrow\uparrow\downarrow\downarrow\uparrow\downarrow\downarrow\downarrow\downarrow}$ and $\ket{\downarrow\uparrow\uparrow\uparrow\downarrow\uparrow\uparrow\downarrow\uparrow\uparrow\uparrow\uparrow}$.  The strategy was to reinforce the orientation of selected qubits (Fig. \ref{fig:ising_cmajor_superposition}), aiming to negate the influence of non-zero $b_{ij}$  coefficients in equation \ref{eq:qubotoising_transform}. This leads to the QUBO coefficients described in \ref{eq:ising_cmaj_superposition}

\begin{figure}[ht]
    \centering
    \begin{tblr}{cccccccccccc}
    % \hline
     \!\!\!\!\!\;\tiny{\textcolor{red}{$-1$}}\:\:\,  & \tiny{$0$}\:\:\, & \tiny{\textcolor{blue}{$1$}}\:\:\, & \tiny{$0$}\!\!\: & \tiny{\textcolor{red}{$-1$}}\:\:\, & \tiny{$0$}\:\:\, & 
     \tiny{$0$}\!\!\: & \tiny{\textcolor{red}{$-1$}}\:\:\, & \tiny{$0$}\;\:\, & \tiny{\textcolor{blue}{$1$}}\;\:\, & \tiny{\textcolor{blue}{$1$}}\;\:\, & \tiny{$0$}\\
    \end{tblr}
    \begin{tblr}{c|[dashed]c|[dashed]c|[dashed]c|[dashed]c|[dashed]c|[dashed]c|[dashed]c|[dashed]c|[dashed]c|[dashed]c|[dashed]c}

        \textcolor{red}{$\uparrow$}\, &  $\downarrow$\, & \textcolor{blue}{$\downarrow$}\, & $\downarrow$\, & \textcolor{red}{$\uparrow$}\, & $\downarrow$\, & $\downarrow$\, & \textcolor{red}{$\uparrow$}\, & $\downarrow$\, & \textcolor{blue}{$\downarrow$}\, & \textcolor{blue}{$\downarrow$}\, & $\downarrow$
    \\

    \end{tblr}

    \begin{tabular}{ccccccccccc}
    % \hline
     \tiny{$>\!\!0$}  & \!\tiny{$<\!\!0$} & \!\tiny{$<\!\!0$} & \!\tiny{$>\!\!0$} & \!\tiny{$>\!\!0$} & \!\tiny{$<\!\!0$} & 
     \!\tiny{$>\!\!0$} & \!\tiny{$>\!\!0$} & \!\tiny{$<\!\!0$} & \!\tiny{$<\!\!0$} & \!\tiny{$<\!\!0$}
     \\
    \end{tabular}
    \caption{Spin configuration of a C Major Chord}
    \label{fig:ising_cmajor_superposition}
\end{figure}

\begin{equation}
\begin{split}
    \begin{aligned}
        b_{kl} = &
\begin{cases}
    1, &\begin{aligned}
        & k \in Chord;\, l \notin Chord \\ & k \notin Chord;\, l \in Chord
    \end{aligned} \\
    -1, & \text{otherwise}
\end{cases}
\\
        a_{k} = &-\frac{1}{2}\sum_{k\neq l}b_{kl} \\
    \end{aligned}  
\end{split}
    \label{eq:ising_cmaj_superposition}
\end{equation}

\begin{figure}[ht!]
  \centering
  \begin{subfigure}{.45\textwidth}
    \centering
    \includegraphics[width=\linewidth]{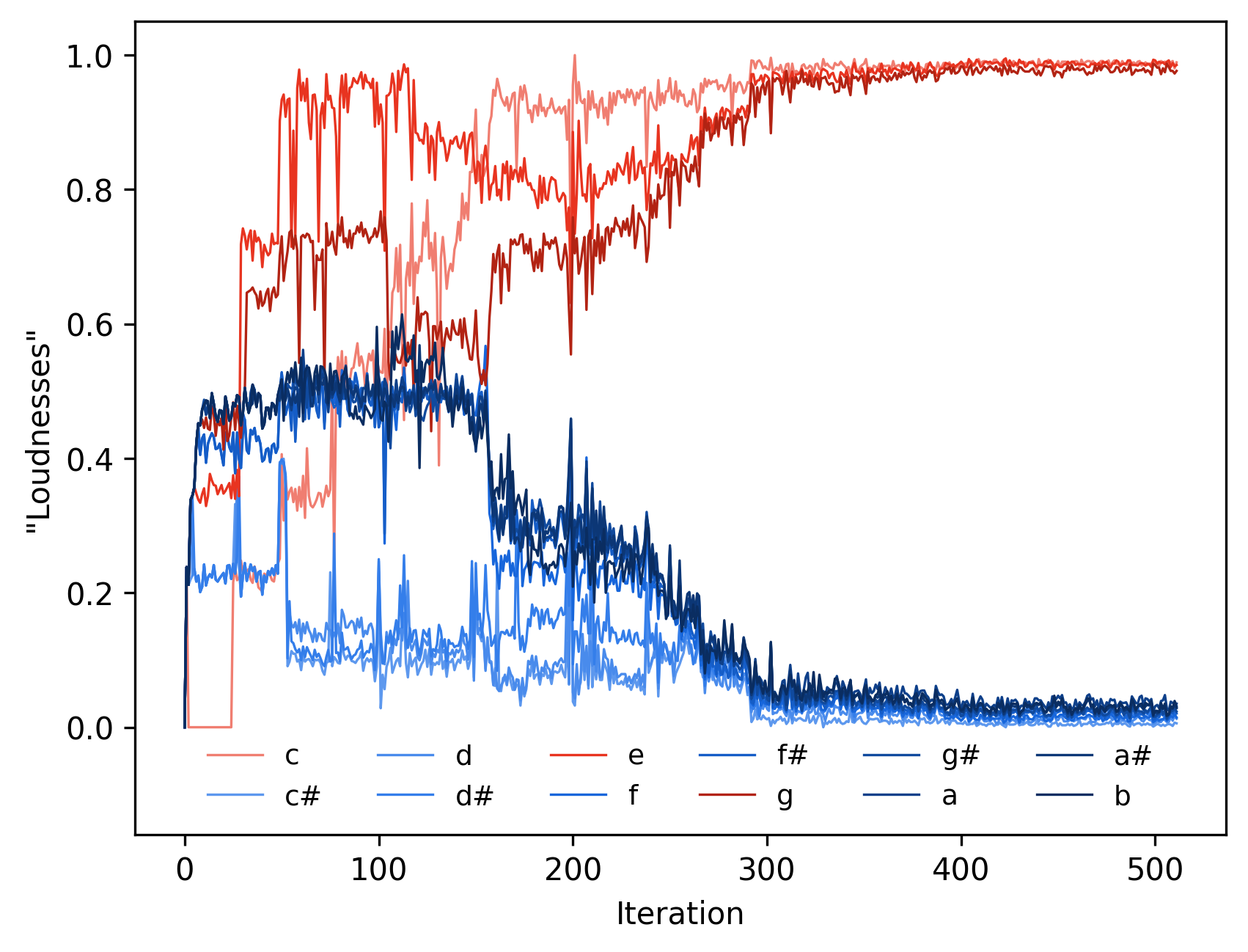}
    \caption{Converging to $C_{maj}$}
    \label{fig:superposition_a}
  \end{subfigure}
  \begin{subfigure}{.45\textwidth}
    \centering
    \includegraphics[width=\linewidth]{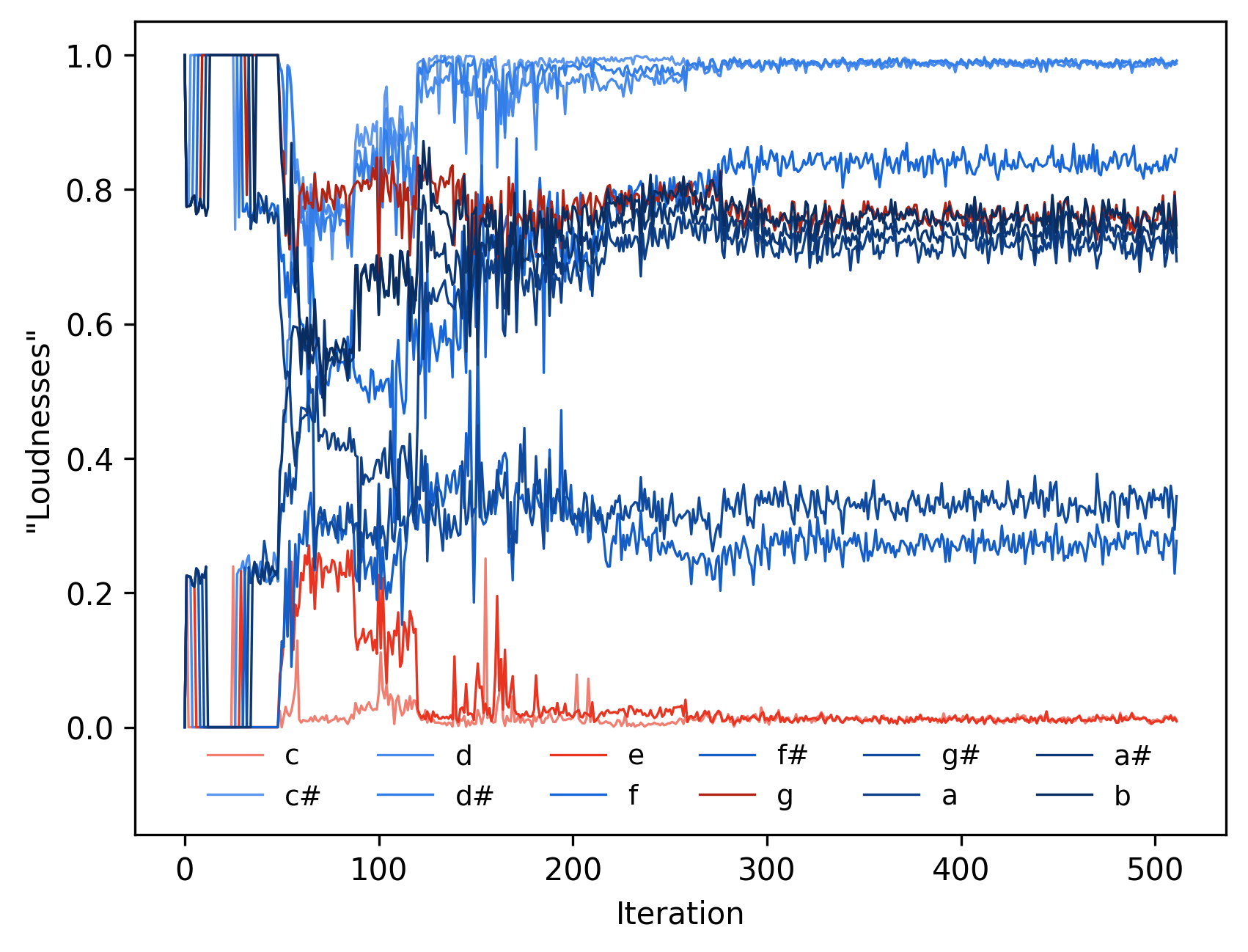}
    \caption{Mixed solution}
    \label{fig:superposition_b}
  \end{subfigure}
  \begin{subfigure}{.45\textwidth}
    \centering
    \includegraphics[width=\linewidth]{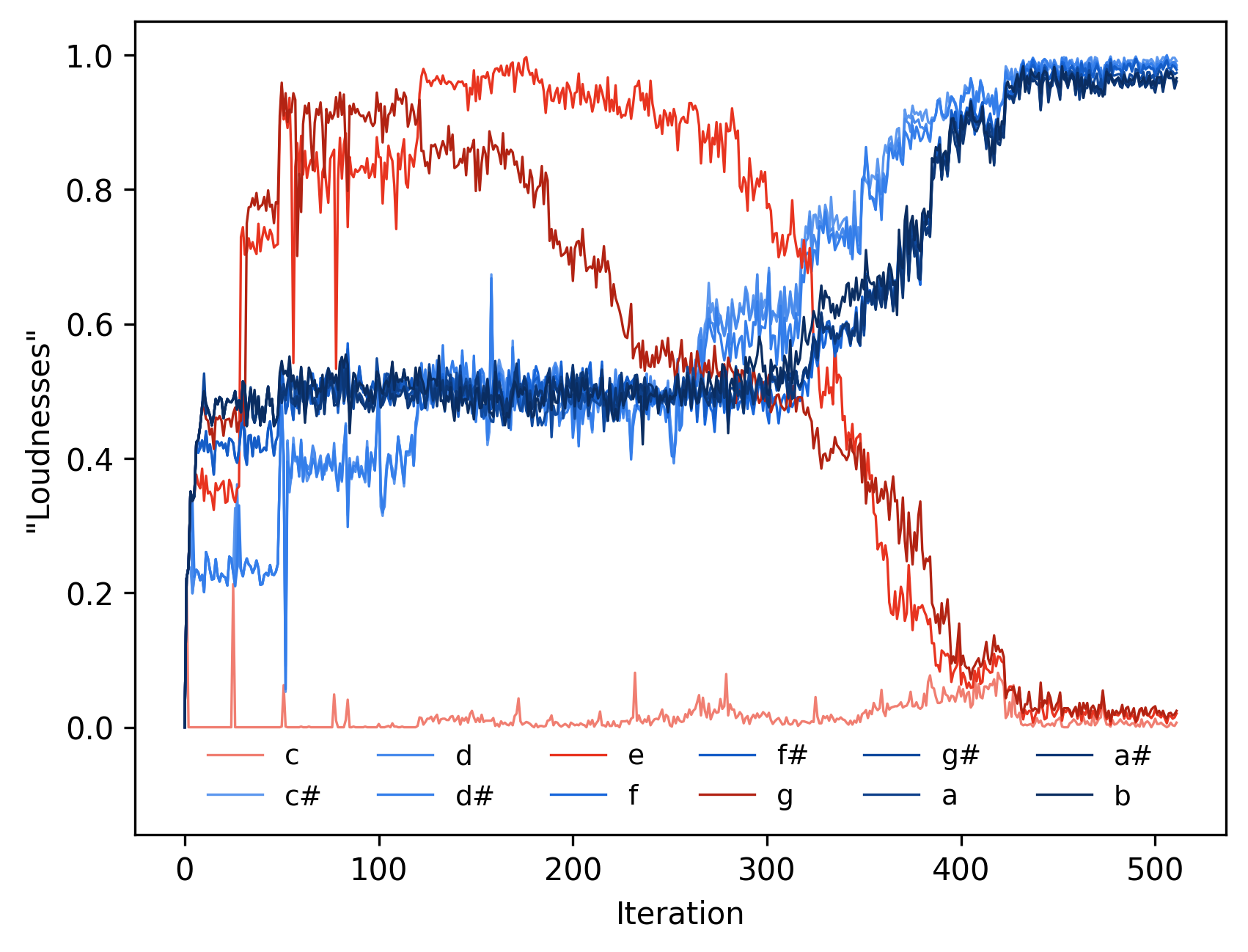}
    \caption{Chord Flip to $antiC_{maj}$}
    \label{fig:superposition_c}
  \end{subfigure}
  \caption{Different VQH results for Example 4}
  \label{fig:superposition}
\end{figure}

Now, it can be verified that $\left\langle H\right\rangle_{C_{maj}} = \left\langle H\right\rangle_{antiC_{maj}} = -12$, converting them into equally attractive results. As a result, different runs of the VQE optimization for this QUBO might arrive in $\ket{C_{maj}}$, $\ket{antiC_{maj}}$, or a mixed state, as shown by Figure \ref{fig:superposition}. 

In the first example, \ref{fig:superposition_a}, it is verified that the $C_{maj}$ chord can be achieved with this QUBO.
In \ref{fig:superposition_b}, a different initial condition was given. The VQH tends to converge to "$antiC_{maj}$"; however, the presence of the 'G' note creates difficulties in the optimization.
Most interestingly, fig. \ref{fig:superposition_c} displays a less common result, where the chord starts approaching a $C_{maj}$, but then completely \textit{flips} the chord after a "stubborn" 'C' note remains unplayed.

\subsection{Adiabatically Navigated VQE}

To improve the efficacy of the VQE algorithm, particularly when searching for the ground state of a quantum system, recent research has proposed the introduction of the adiabatic theorem for navigating towards the solution. The method proposes to have a time-dependent Hamiltonian that is gradually (and slowly) moving towards an asymtoptic equilibrium. For instance, consider a system that starts at an eigenstate $\psi_0$ of a given Hamiltonian $H_0$ . The system evolves in time and is subjected to an adiabatic process, until it reaches a final eigenstate $\psi_1$ of a different Hamiltonian, $H_1$. 

Matssura et.al. \cite{Matsuura_2020} proposes a method of an adiabatically navigated quantum eigensolver, which involves a time-dependent Hamiltonian using the same approach adopted in the Example 2. A simple approximation of an adiabatic process (eq. \ref{eq:adiabatic}) was used to \textit{smooth} the transition between different VQE. From an initial Hamiltonian $H_{initial}$, VQE is used to find it's eigenstate - which becomes the initial condition for the next VQE. Then, the time $t$ is updated, leading to a slightly different configuration. Eventually, after $m$ updates, the system arrives at $H_{final}$ \textit{adiabatically}.

\begin{equation}
    H(t) = (1-t)H_{initial} + tH_{final}
    \label{eq:adiabatic}
\end{equation}

\subsection{Example 5: Adiabatic Chord Transitions}
\begin{figure}[ht!]
    \centering
    \includegraphics[width=.45\textwidth]{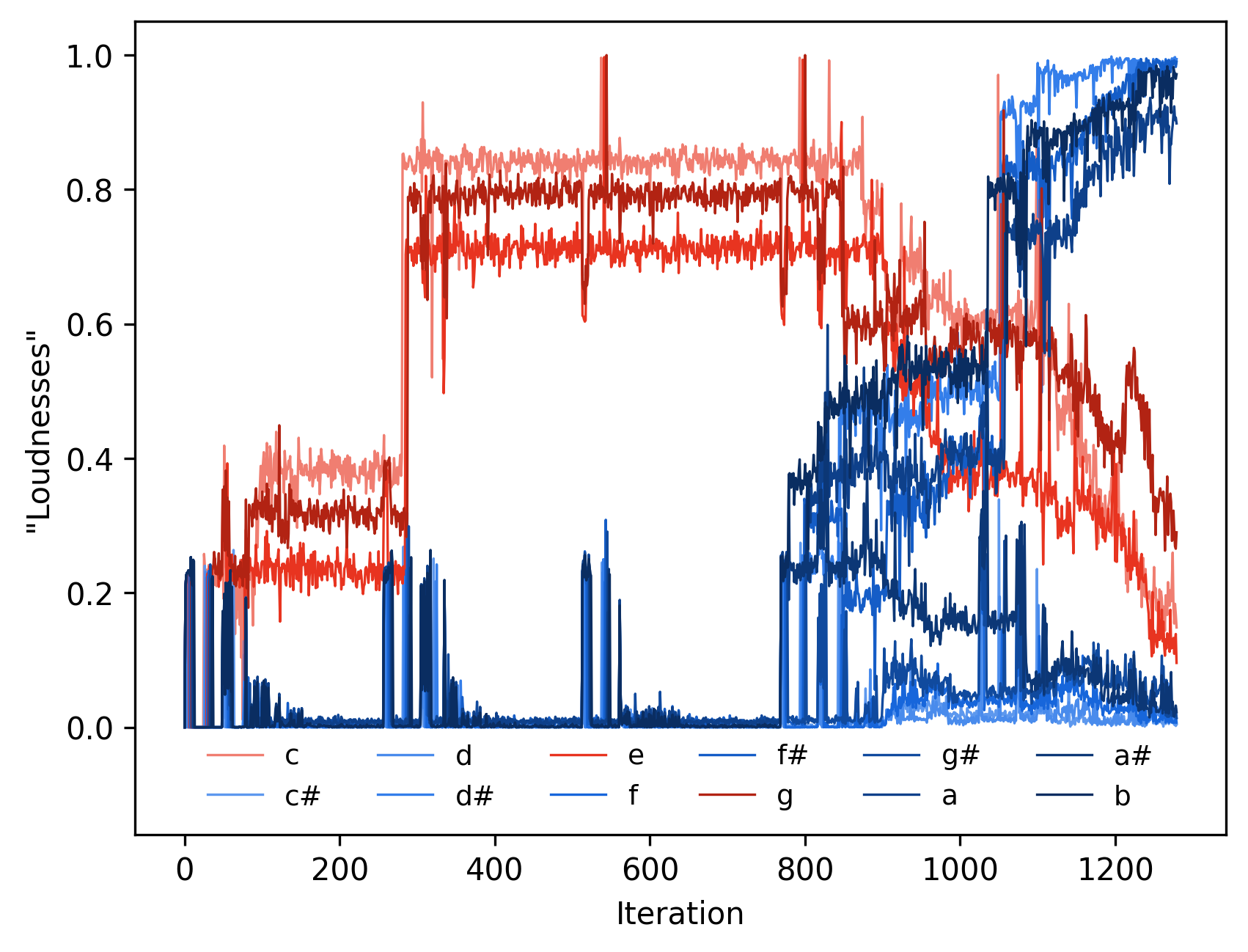}
    \caption{Example 5: Adiabatic Chord Progression}
    \label{fig:adiabatic}
\end{figure}

The same principle can be applied to chord progressions. Instead of making sharp perturbations and transitions to different chords (as in Example 2), one could start playing an "Ising Harp" at a $\ket{C_{maj}}$ configuration, and make an \textit{adiabatic chord transition} to, say, $\ket{B7/D\#}$ (Fig. \ref{fig:adiabatic}). For simplicity, both chords used a linear QUBO function (as in Example 1).

\section{Using the VQH as a Musical Interface}
\label{sec:interface}
For a scientifically driven approach to sonification, the above investigation provides more insight about the inner workings of the VQE algorithm, as well as more intuition about how changing specific parameters leads to different results. The next question arises: How to use the VQE \textit{musically}?

The current VQH implementation is presented as a CLI software implemented in Python, that is used to both run VQE simulations with user-provided specifications and also redirect the results to a synthesis engine (in this case, SuperCollider) for sonification. The user can interface with the system by changing the VQE specifications (e.g., number of iterations, classical optimizer, amount of hamiltonians for chord progressions, etc.) and updating QUBO coefficients. This can be done by either using a text editor, or a mapped MIDI controller.

\subsection{Interfacing with a text editor}
\begin{figure}[ht!]
    \centering
    \begin{subfigure}{.45\textwidth}
    \centering
    \includegraphics[width=\linewidth]{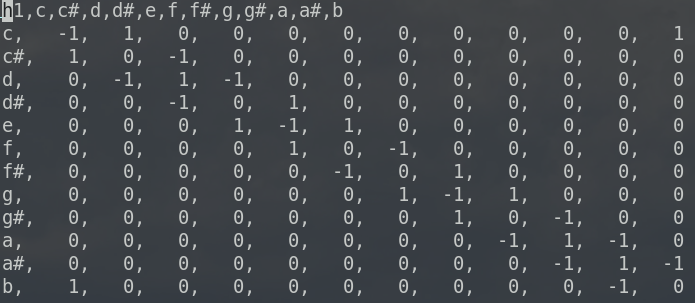}
    \caption{QUBO Coefficients for Example 4}
    \label{fig:h_setup}
  \end{subfigure}
  \begin{subfigure}{.45\textwidth}
    \centering
    \includegraphics[width=0.8\linewidth]{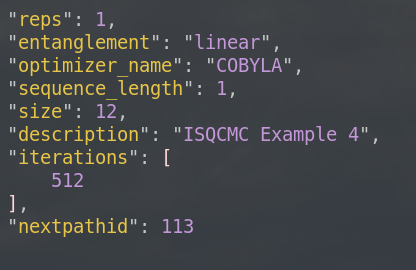}
    \caption{VQE Parameters}
    \label{fig:vqe_conf}
  \end{subfigure}
    \caption{Text-Based VQH Interface}
    \label{fig:vqh_setup}
\end{figure}

The QUBO coefficients are stored and read from a \texttt{.csv} file in matrix form, as depicted in Fig. \ref{fig:h_setup}.  The first row gives labels for the notes, providing a namespace that is used in the sonification mapping. The diagonal terms correspond to the linear terms $a_i$ of eq.\ref{eq:qubo}, whereas the nondiagonal terms define couplings between notes.
A separate \texttt{.json} file contains other relevant VQE configuration, such as the number of iterations.

After updating and saving the changes made in both files, the user can use the \texttt{runvqe} function at the VQH prompt to generate new data, followed by the \texttt{play} function to trigger a sonification.

This workflow is specially appealing for \textit{Live Coding} performances, where artists implement or change software configurations on stage, leading to on-the-fly changes on the music. The artist's screens and text editors are commonly displayed to the audience, where they can track the changes being made, and how they affect the resulting sounds.

\subsubsection{Example 6: Obtaining Musical Variations by Changing the Classical Optimizer}
\begin{figure}[ht!]
\centering
  \begin{subfigure}{.45\textwidth}
    \centering
    \includegraphics[width=\linewidth]{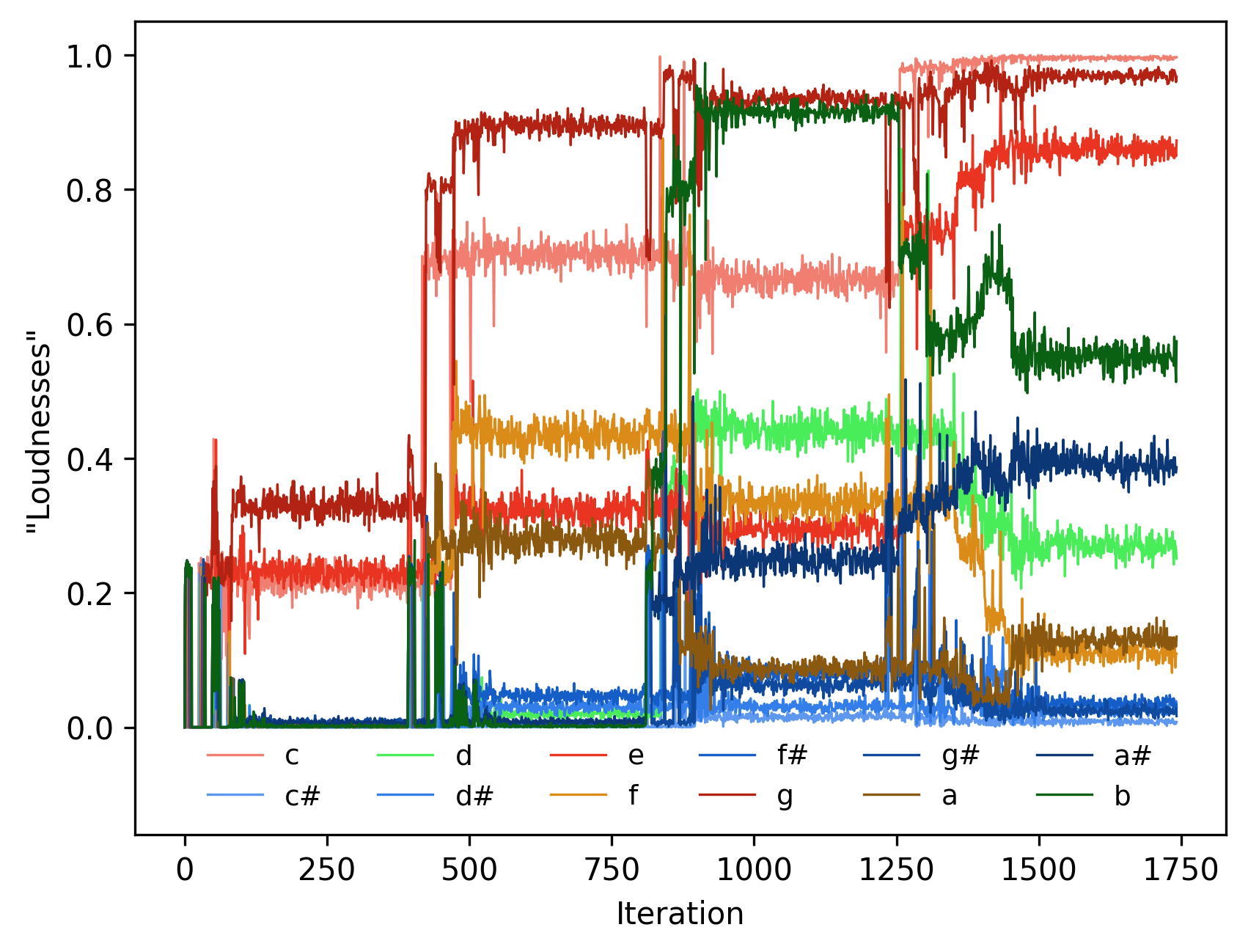}
    \caption{COBYLA}
    \label{fig:cobyla}
  \end{subfigure}
  \begin{subfigure}{.45\textwidth}
    \centering
    \includegraphics[width=\linewidth]{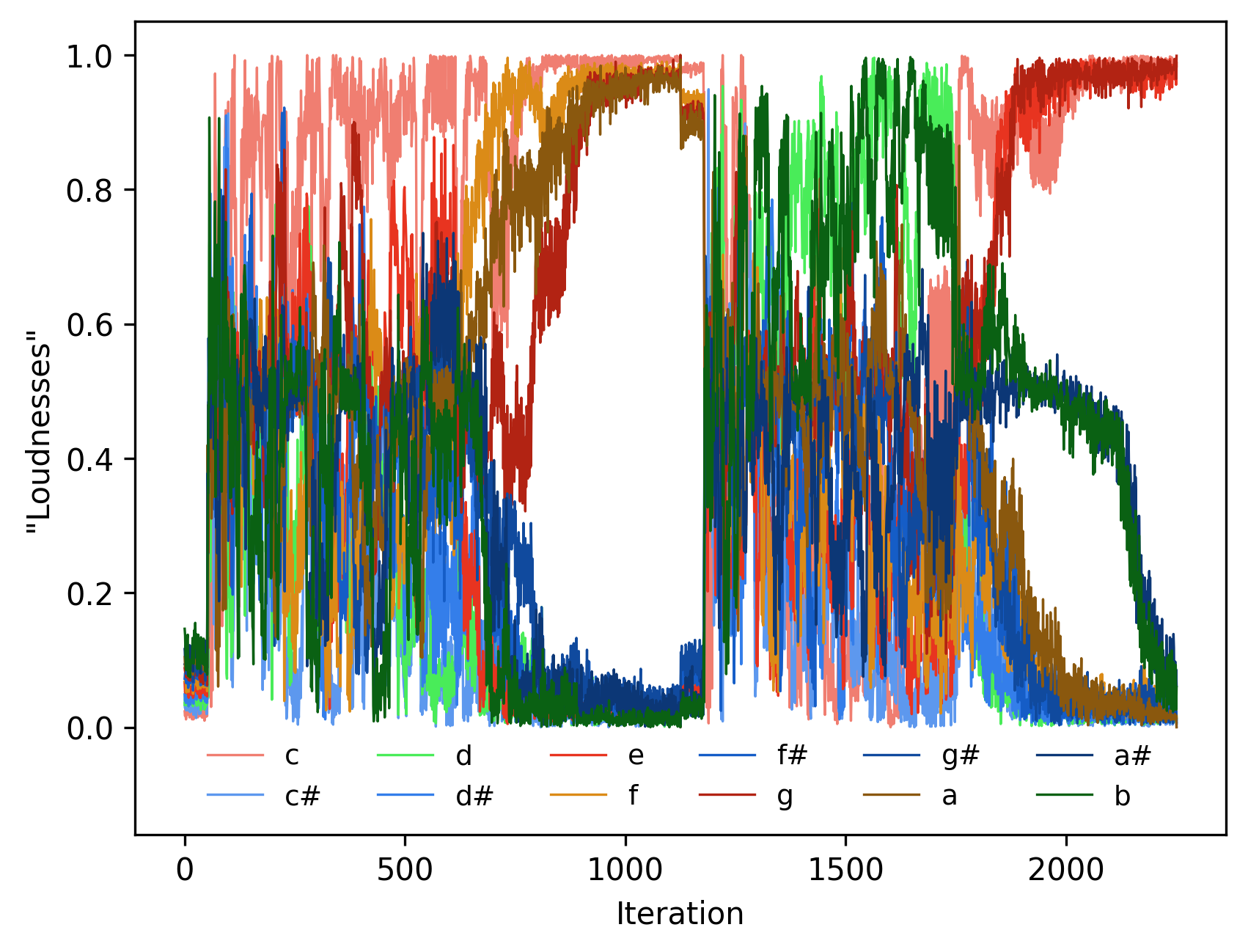}
    \caption{SPSA}
    \label{fig:spsa}
  \end{subfigure}
  \begin{subfigure}{.45\textwidth}
    \centering
    \includegraphics[width=\linewidth]{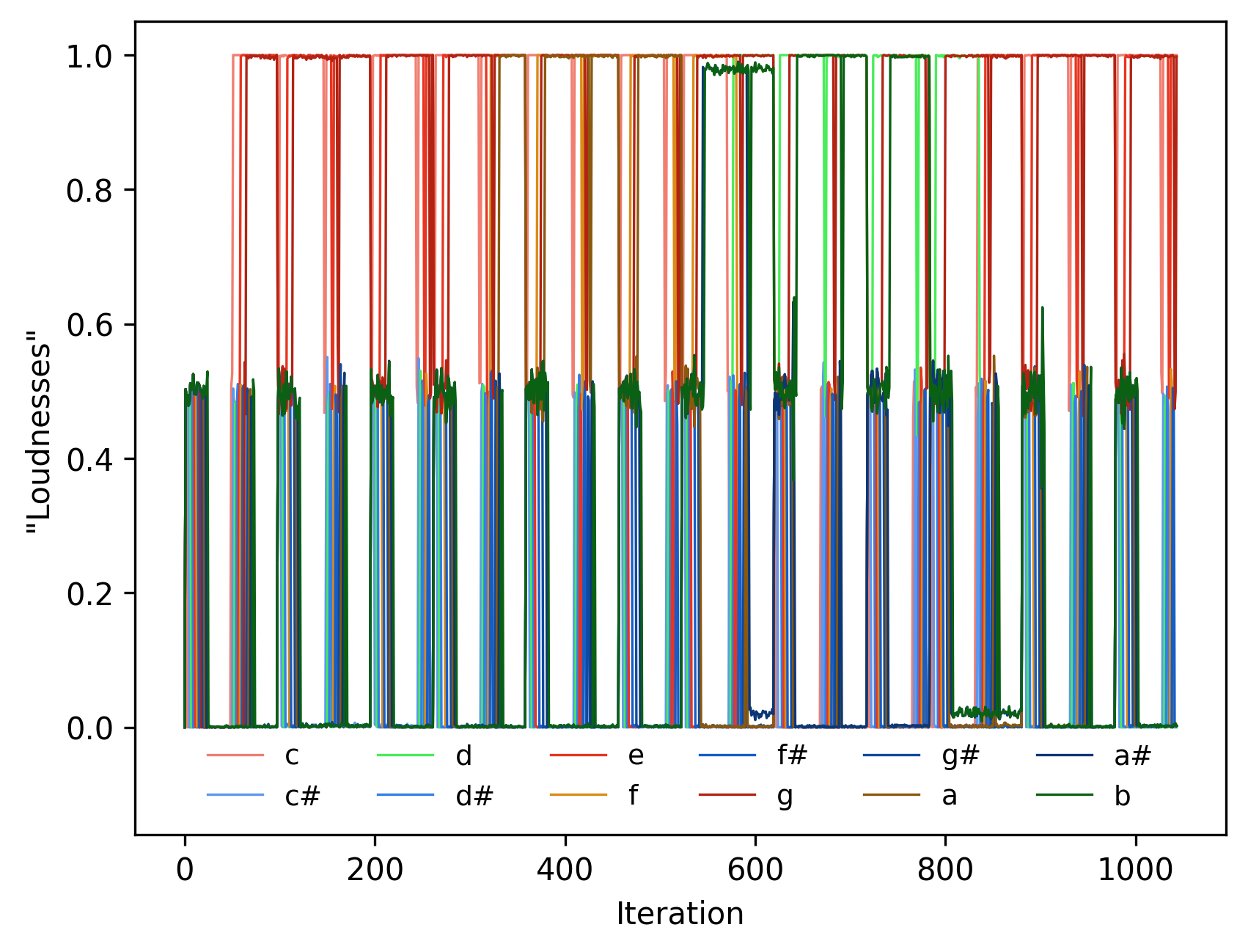}
    \caption{NFT}
    \label{fig:nft}
  \end{subfigure}
    \caption{Comparison between Optimizers}
    \label{fig:optimizers}
\end{figure}

Notice how the music can be radicallly changed by simply switching between classical optimizers. In Fig. \ref{fig:optimizers}, the QUBO from Example 1 was optimized using different techniques. In this work, COBYLA (\ref{fig:cobyla}) seems to provide eventual \textit{note sweeps} and persistent mild attacks on all notes, mimicking a "drum roll". SPSA (\ref{fig:spsa}), on the other hand, takes longer (or struggles) to converge, meaning that it usually onsets and keeps all the notes sounding simultaneously, until penalised notes progressively fade out. In contrast, the NFT algorithm (\ref{fig:nft}) provides a periodic, rythmic pattern that can be exploited.

\subsection{Using a MIDI interface to Manipulate QUBOs}

\begin{figure}[ht!]
    \centering
    \includegraphics[width=.35\textwidth]{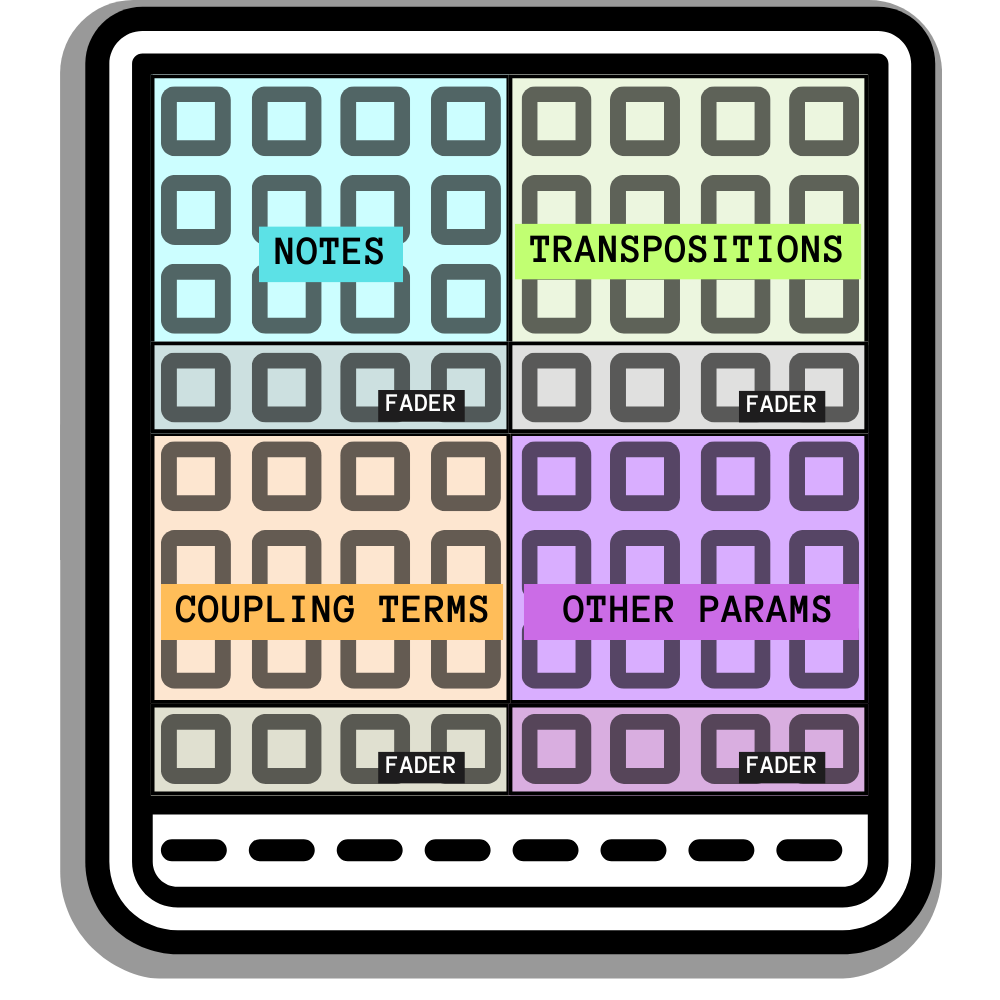}
    \caption{Launchpad X MIDI Interface Design}
    \label{fig:interface}
\end{figure}

A possible approach for an interface to control the QUBO matrix through a MIDI device is now presented. The proposal consists of an eight by eight grid of buttons divided into four different sections. A simple scheme of the interface is shown in Fig. \ref{fig:interface}. A device that provides this structure is, for example, is the Launchpad X.

When a button is pressed, a MIDI note is sent to the main program for further processing. The interface incorporates a visual feedback mechanism, utilizing lights, to inform the user about the information being transmitted and the activated buttons. Each section of the grid serves a different purpose. The top left four by four grid is allocated for the diagonal terms of the QUBO matrix, which correspond to the 12 notes. Each button within this section functions as a toggle, allowing the spin value to vary between 1 and -1. When a note is activated, its corresponding button is pressed, and its value changes from -1 to 1. It is worth noting that the last row of this section is distinct, as the grid size is four by four, while there are only 12 notes. This last row is designated as a discrete fader, and the same approach is employed for the last row of the remaining three sections. These faders are essential for the bottom left section, which is responsible for handling the coupling terms of the second approach of the QUBO matrix. Since there are only 11 coupling terms, whereas there are 12 buttons in this section, the first and last spins of the one-dimensional Ising model can be interconnected to include an additional coupling term. In this section, if a button is held and the user slides through the faders, the value of the coupling term associated with that button can be adjusted. The range and discrete values for the coupling values must be specified.

The top right corner of the grid is utilized for transpositions. Specifically, four buttons are allocated to apply octave and semitone transpositions, allowing both pitch increases and decreases. The remaining eight buttons in this section can be used for implementing other transpositions, such as fourth and fifth transpositions, since these intervals are also generators of the equal temperament's structure. Finally, the bottom right corner is left unassigned, paniding room for additional controllers dedicated to other parameters, such as an external magnetic field (see Sec.\ref{sec:future_work}).

\section{Sonification Mapping Strategies}
\label{sec:mapping}
After exploring the VQH interface with potential design strategies for musical applications, it is time to investigate different mapping strategies that can lead to more complex and abstract musical results. As mentioned, this is known as a \textit{mapping problem} in the sonification literature. The data being sonified, as explained in section \ref{sec:dataset}, are streams of marginal distribution coefficients with a respective expectation value.

The initial ideas for mapping methods for these data are discussed in this section. The chord progression from Example 2 will be used as a guideline. The synthesizers for each approach mentioned in this work were implemented in SuperCollider.

\subsection{Additive Synthesis}
There is a natural additive framework that arises when interpreting qubits as individual notes, as discussed in section \ref{sec:qubit_chords}. For the 12-qubit examples shown so far, there would be twelve oscillators with well-defined frequencies (such as the chromatic scale). The Marginal distribution coefficients are then used as the amplitudes of each oscillator. Oscillators could contain custom frequencies and waveforms (such as sawtooth, square, or custom samples).
\begin{figure}[ht!]
    \centering
    \includegraphics[width=.45\textwidth]{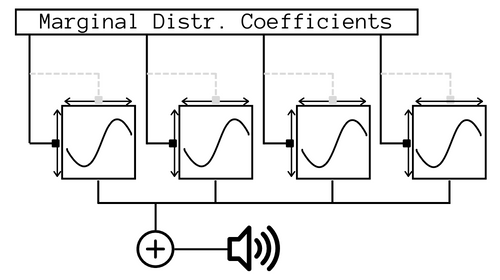}
    \caption{Additive Synthesis Mapping}
    \label{fig:additive}
\end{figure}
\subsubsection{Inharmonicity Approach}

Another additive aproach to statevector sonification focuses on modulating frequency, instead of amplitude. For example, starting from a 12-note harmonic series, it is possible to define a proportional \textit{shift} $c_n(t)$ of a given harmonic term (eq. \ref{eq:inharmonic}). This introduces dynamic inharmonicities that can be explored musically.

\begin{equation}
    f_n(t) = (n - c_n(t))f_{1}
    \label{eq:inharmonic}
\end{equation}

\subsection{Subtractive Synthesis}
To improve the sonification timbre, a possible follow-up approach is usually to "broaden up" the frequency bands of the individual oscillators. A practical solution for this is to invert the synthesis and use a subtractive or filtering approach instead. In simple terms, a white noise is generated and then filtered through narrow bandpass filters centered at the desired frequencies. Additionally, the expectation values can be used to control the quality factor of the band-pass filters, in a way that if the estimated energy decreases, the filter narrows down. In other words, the farther away the VQE is from the exact result, the noisier the sound gets.
\begin{figure}[ht!]
    \centering
    \includegraphics[width=0.45\textwidth]{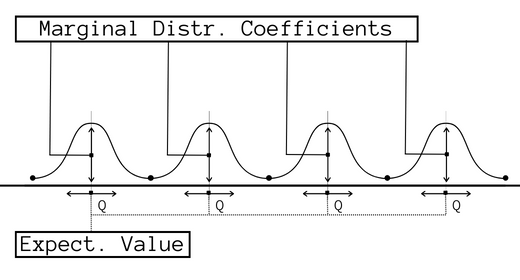}
    \caption{Subtractive Synthesis Mapping}
    \label{fig:subtractive}
\end{figure}
\subsection{Arpeggios}
Arpeggiation is a mapping strategy that can be utilized to enhance the perception of dynamics using a temporal expansion. The marginal distributions are again mapped as amplitudes.  However, at each iteration, the notes are expanded into an arpeggio and sorted by amplitude. In other words, the most intense note is played last in the arpeggio. Furthermore, the expectation values can be assigned as the \textit{expansion rate} of the arpeggio. In other words, the closer from the exact result, the faster and denser the arpeggio will be.
\begin{figure}[ht!]
    \centering
    \includegraphics[width=.45\textwidth]{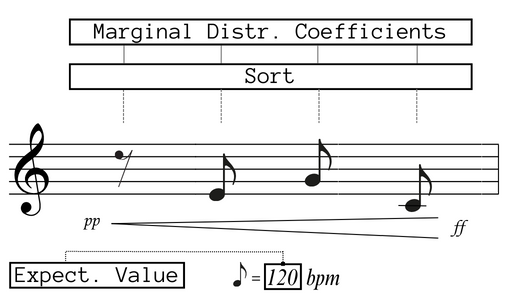}
    \caption{Arpeggio Mapping}
    \label{fig:arpeggio}
\end{figure}

\section{Using the VQH in live performances and composition}
\label{sec:live}

\subsection{Dependent Origination: Dynamically changing the sonification mapping on stage}
Dependent Origination is a piece composed by Peter Thomas and Paulo Vitor Itaboraí. It has been performed twice to date, at the IKLECTIK Arts Lab in London\cite{IKLECTIK} and ICFO in Barcelona (see institutional video\cite{ICFO}. The VQH interface appears at \texttt{0:12}) - using Zen, a live coding application designed and built by Thomas\cite{Zen}. In this section, we will briefly describe the Zen system, outline the aesthetic motivations behind the composition, followed by a more technical explanation of how the VQH was used to generate data for sonification.

\begin{figure}[ht!]
    \centering
    \includegraphics[width=0.45\textwidth]{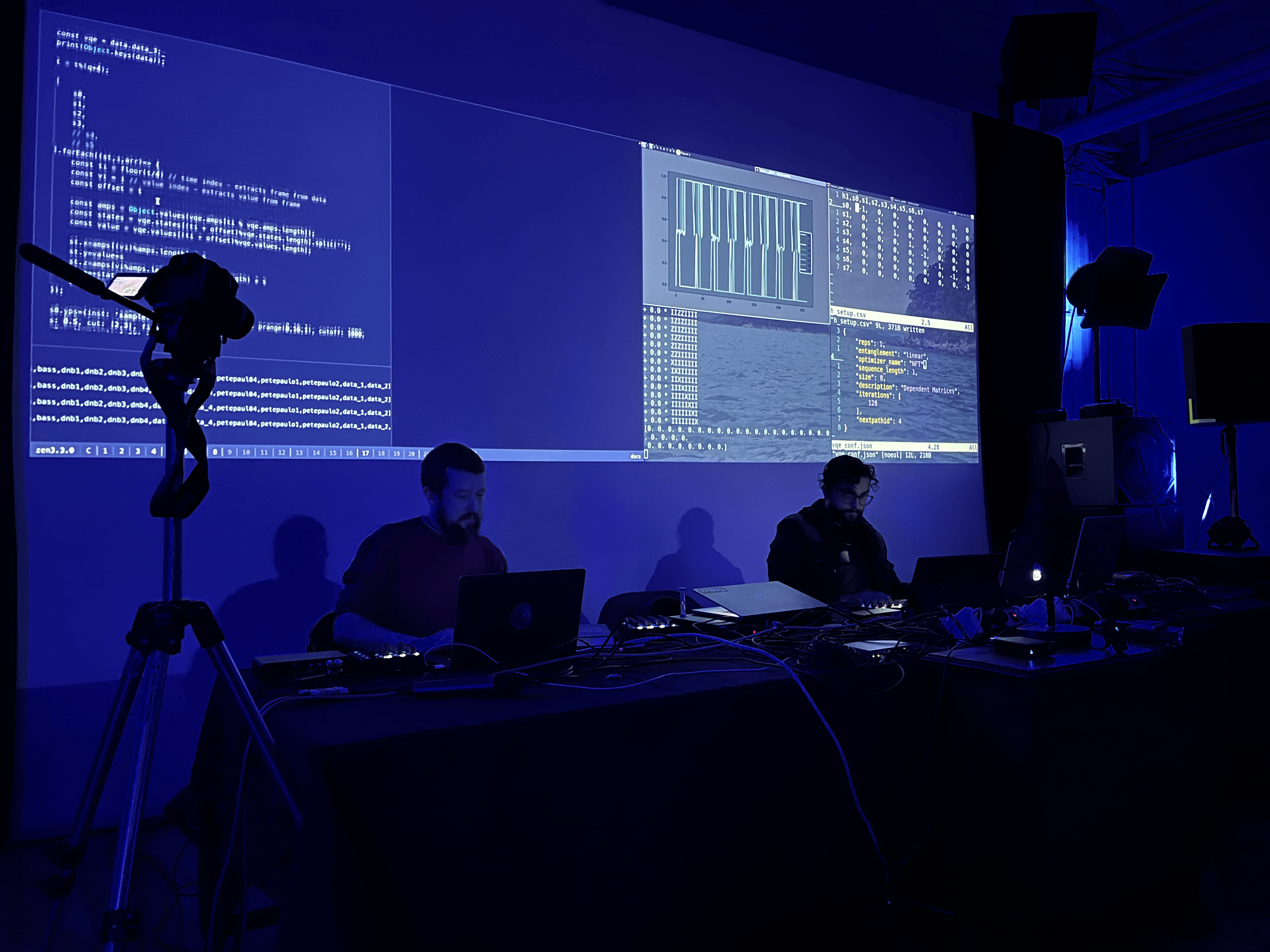}
    \caption{Performance of \textit{Dependent Origination} at IKLECTIK}
    \label{fig:enter-label}
\end{figure}

Zen is a JavaScript library for expressing multidimensional, musical patterns using single line expressions, with a particular emphasis on pattern interference. In addition to the language, the Zen ecosystem is further comprised of an integrated development environment (IDE) - including a code editor, pattern visualiser and synthesis engine (Oto). Each component is built for the web, resulting in a holistic performance tool that requires no installation beyond a modern browser. 

Through the textual interface, Zen allows the user to succinctly express causal relationships between musical layers. A composer is able to define discrete patterns, then identify musical or sonic parameters between patterns that should interfere with each other; for example, pitch, amplitude, timbral or spatial parameters. Composing in this way leads to surprising, often unintended results, challenging the common portrayal of inspiration as a form of guiding agent; instead, seeking to stimulate ideas through processes that go beyond the mind, as a means of creating original work; see Fell \cite[p.~21]{fell2022structure} and Magnusson and McLean \cite[p.~262]{magnussonmclean2018} for further exposition upon the value of harnessing processes that go beyond the imagination of the composer.

Dependent Origination is a pillar of Buddhist philosophy that emphasises the interrelation of all phenomena. Everything exists in dependence upon its particular causes and conditions. Everything is impermanent, and no phenomena arise from nothing. This interpenetration of phenomena has poetic echoes in quantum entanglement; a fundamental aspect of quantum mechanics and quantum computing. Entangled particles may affect each other even when separated by long distances, even without mechanical contact. In addition to expressing causal relationships between musical layers within the Zen code, we therefore also sought to explore pattern inference within the design of the VQH interface.

Zen allows the user to map musical and sonic parameters across a three-dimensional canvas. Eight separate streams can be moved independently around the space, with the parameters of sonic events being determined by their current position in time and space. In an eight-qubit system, we mapped the marginal coefficients to the XY-positions of the separate streams on the Zen canvas. Each stream was assigned to a separate instrument; which included drum samplers and FM and granular synthesizers. Sonic events were triggered using the 8-bit binary string returned at each iteration, assigning one bit to each stream and triggering an event when the value was equal to 1. The resulting rhythms and spatial positioning was used as the basis for improvisation in real-time, by assigning different sonic parameters to each axis to provide variety, offsetting layers and swapping out the underlying quantum data as the composition progressed.

\section{Concluding Thoughts}
\label{sec:conclusion}
Sonification, as a cross-disciplinary practice, is a fertile ground for exploring and combining different approaches, methodologies and interpretations for the emerging quantum computer field. 

Artists, driven by creative expression and aesthetic exploration, approach sonification as a means to evoke emotional responses and convey subjective experiences. They prioritize the artistic \textit{interpretation} of the data, focusing on the qualities of the resulting soundscapes. Artists often employ metaphorical or abstract representations, manipulating and transforming data to create unique sonic experiences that engage listeners on a visceral and emotive level. Their aim is to provoke thought, inspire contemplation, or challenge established perceptions of the world. In other words, the data is incorporated into a musical instrument.

In contrast, scientists approach sonification with a primary focus on data analysis, comprehension, and scientific discovery. They aim to enhance the understanding of complex data sets by utilizing sound as an additional modality for conveying information. Scientists tend to prioritize fidelity and accuracy in sonification, striving to maintain a direct and transparent mapping between the data and the resulting auditory representation. 

This work has attempted to collide both approaches into a resulting Musical Interface, that can both be a tool for enhancing data visualization and creating artistic pieces.

\subsection{Future Work Considerations}
\label{sec:future_work}
There are a few paths in which the authors envision that this work could unfold. Initially, the QUBO function is applied in a large set of problems, and researchers might take interest on exploring sonifications of their own particular problems using QUBO and VQH. Secondly, the Ising problem used in this work considered only fields (linear coefficients) pointing in the same direction of the spin lattice ($Z_i$), which effectively behaves as a classical model. However, quantum behaviour start to appear when a Transverse Magnetic Field is applied (eq. \ref{eq:isingtransverse}), (as in Clemente's et.al. work \cite{clemente2022new}), leading spins to be in states such as $\ket{\leftarrow}$ and $\ket{\rightarrow}$, which are inherently quantum. Additionally, the intensity of the field $h_x$ will dictate the magnetization phase of the system. As a result, $h_x$ becomes a potential control parameter for the VQH instrument.
\begin{equation}
H(\sigma)=\sum_i^N h_x \sigma^X_i + \sum_i^N\sum_{j<i}^N b_{ij}\sigma^Z_i \sigma^Z_j
\label{eq:isingtransverse}
\end{equation}

Furthermore, quantum circuit design of the variational form (sec. \ref{sec:circuits}) will likely provoke a significant impact on the optimization process and produce completely different sounds, as consequential as changing the classcal optimizers. A thorough sonification comparison between a larger collection of classical optimizers and ansatze could provide important insights.

Additionally, from a mapping perspective, researchers could propose new encoding strategies of the QUBO problem (different from the \textit{one-note-per-qubit} approach taken), and a myriad of sonification/musification strategies for the data generated.

Finally, the VQH interface itself can be improved (source code available on GitHub at the time of publication \cite{VQHGit}) to allow new music expressivity and quantum-computer-assisted composition.

\section{Acknowledgements}

We would like to thank Y. Chai and S. K\"uhn for fruitful discussions. We also thank the IKLECTIK Arts Lab for programming the concert in which \textit{Dependent Origination} was premiered to the general public.

K.J. and A.C.\ are supported with funds from the Ministry of Science, Research and Culture of the State of Brandenburg within the Centre for Quantum Technologies and Applications (CQTA). 
\begin{center}
    \includegraphics[width=0.1\textwidth]{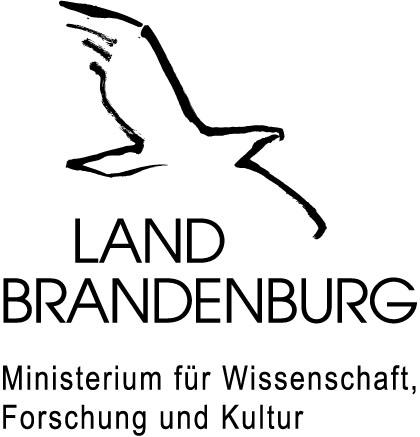}
\end{center}
K.J.'s work is funded by the European Union’s Horizon Europe Framework Programme (HORIZON) under the ERA Chair scheme with grant agreement no.\ 101087126.

\bibliographystyle{unsrt}
\bibliography{main}

\end{document}